\documentclass{aa}  
\pdfoutput=1
\usepackage[draft]{pgf}
\usepackage{graphicx}
\usepackage{rotating}
\usepackage{amsmath}
\usepackage{amssymb}
\usepackage{txfonts}
\usepackage{color}
\usepackage{gensymb}
\usepackage{soul}
\usepackage{txfonts}
\usepackage{color} 
\usepackage{soul}
\usepackage{booktabs}
\usepackage{multirow}
\usepackage{mathtools}
\def\teff{T_{\text{eff}}}
\def\stage{t_{\star}}
\def\stmass{M_{\star}}
\def\msol{M$_{\odot}$}
\def\alphaov{\alpha_{\mathrm{ov}}}
\def\xv{\boldsymbol{x}}

\def\epsilonv{\boldsymbol{\epsilon}}
\def\thetav{\boldsymbol{\theta}}
\def\thetastv{\boldsymbol{\theta_{\star}}}
\def\Sv{\boldsymbol{\mathcal{S}}}

\def\Sigmav{\boldsymbol{\Sigma}}

\def\rzo{r_{01}}
\def\rzoz{r_{010}}
\def\rzt{r_{02}}
\def\rot{r_{13}}

\begin{document}

   \title{Uncertainties and biases in modelling 16 Cyg A and B}


   \author{
     M. Bazot\inst{1,2}
   }

   \institute{Division of Sciences, New York University Abu Dhabi, United Arab Emirates\email{mb6215@nyu.edu}
     \and
    Center for Space Science, NYUAD Institute, Abu Dhabi, UAE
             }

   \date{Received September , ; accepted March , }

 
  \abstract
   {} 
   {In this study I assess how existing data for the solar analogues 16 Cyg A and B, in particular the asteroseismic measurements obtained from \emph{Kepler}, constrain theoretical stellar models. The goal is two-fold: first to use these stars as benchmarks to discuss  which precisions can realistically be expected on the inferred stellar quantities; and second to determine how well {`\lq}non-standard'{\rq} prescriptions, such as microscopic diffusion and overshoot, are constrained.}
   {I used a Bayesian statistical model to infer the values of the stellar parameters of 16 Cyg A and B. I\ sampled the posterior density of the stellar parameters via a Markov chain Monte Carlo (MCMC) algorithm, tested different physical prescription, and examined the impact of using different seismic diagnostics.}
   {General good agreement is found with several recent modelling studies on these stars, even though some discrepancies subsist regarding the precise estimates of the uncertainties on the parameters. An age of $6.88\pm0.12$~Gyr is estimated for the binary system. The inferred masses, $1.07\pm0.02$~ for  Cyg A and $1.05\pm0.02$~ for  Cyg B, are shown to be stable with respect to changes in the physical prescriptions considered for the modelling. For both stars, microscopic diffusion has a significant effect on the estimates of the initial metallicity. Overshoot is confined to very small regions below the convective zone. I show that a proper treatment of the seismic constraints is necessary to avoid  biases in the estimate of the mass.}
   {}


   \maketitle
%

\section{Introduction}

In recent years, asteroseismology has become one of the most important tools to obtain precise estimates of stellar parameters. The first conclusive results for Sun-like stars were obtained from ground-based telescopes using high-precision spectrographs. The following bright stars were first observed: $\alpha$~Cen~A \citep{Bouchy01,Bedding01,Bazot07}, $\alpha$~Cen~B \citep{Bedding05}, $\mu$~Ara \citep{Bouchy05}, $\iota$~Hor \citep{Vauclair08}, $\beta$~Hyi \citep{Bedding07}, and 18 Sco \citep{Bazot12}. The space missions CoRoT \citep{Baglin09} and \emph{Kepler} \citep{Borucki10} later provided much longer photometric time series for many more fainter stars. In particular, \emph{Kepler} allowed for the study of an important number of Sun-like stars and the ability to obtain precise estimates of their masses, ages, and initial chemical compositions \citep{SA17}.

Among the Sun-like stars observed by Kepler, the solar analogues 16~Cyg A and B stand out as particularly important. These are two of the brightest stars observed during the mission. Very precise measurements of their photometric flux were obtained over a period of 2.5 years. Spectral analysis of the resulting flux time series resulted in the detection of 54 and 56 global pulsation modes for 16 Cyg A and B, respectively \citep{Metcalfe12,Davies15}. For both stars, the uncertainties on the estimated eigenfrequencies are of the order of a few tenths of microhertz. Such precisions have allowed us to constrain stellar models tightly and obtain in turn a very good precision on the physical parameters of these stars, most notably their ages. \citet{Metcalfe15} claim estimated uncertainties on the age of the order of 250~Myr. Modelled separately, 16 Cyg A and B appear to have ages 7.07~Gyr and  6.74~Gyr, respectively, favouring an age for the system around 7~Gyr. Based on these studies, the masses of these stars are expected to be slightly higher than the Sun, by a few percent, and their initial metallicities almost solar. These results confirmed previous spectroscopic studies that classified these stars, based on their atmospheric characteristics, as good solar analogues and sometimes even as solar twins \citep[for a discussion, see for instance][]{PdM14}.

It has often been argued that solar twins and analogues are good candidates to test non-standard stellar physics \citep[see e.g.][]{Bazot18}. This is because current stellar models have been mostly calibrated using the Sun, relying on the very good precisions of solar data. Consequently, they are usually well tested in regimes close to the solar regime, even for {}non-standard{} physics such as overshooting or microscopic diffusion \citep{Basu16}. The interest in using solar twins and/or analogues is thus that they can be modelled using the same physics without making further assumptions. This implies that we can confidently test these non-standard physics for stars that are not the Sun and therefore explore their behaviour when stellar physical characteristics, such as mass, age, or chemical composition, vary.

Another important aspect of studying 16 Cyg A and B is that the exquisite precision on their frequencies allow us to discuss the statistical reliability of the estimates of the physical parameters of 16 Cyg A and B. This is critical since observable stellar quantities, such as effective temperature, luminosity, seismic frequencies, radius, and metallicity, depend non-linearly on the stellar parameters of mass, age, initial chemical composition, and mixing-length parameter. Therefore good precisions on the data are needed to avoid complex behaviours of the underlying densities of the stellar parameters \citep[e.g.][]{Bazot18}. Even though some estimates of the stellar parameters may be found to vary significantly in some studies because of the variety of estimation strategies and numerical codes used for the modelling, 
most of these estimates show remarkable agreement. For instance, the estimated ages of 16 Cyg A span a range of $\sim$0.9 Gyr in \citet{SA17}, for an average value of the error bars of the order of 0.35 Gyr. In the framework of stellar-parameter estimation, the cases of 16 Cyg A and B the probability densities of the parameters are strongly constrained by the data. This allows for a careful discussion not only concerning the agreement of various estimates of their physical quantities, but also regarding the robustness of the associated uncertainties.

\begin{table*}
\center
\caption{Non-seismic observational properties of 16~Cyg~A and B.}
\label{tab:obs}      
\centering                                      
\begin{tabular}{lcccc}
\hline\hline
\\[-3.mm]
Star &$\teff$ (K)& [Fe/H] & $L/L_{\odot}$ & $R/R_{\odot}$ \\
\hline
\\[-3.mm]
16 Cyg A & $5825\pm 50$  & $0.10\pm0.09$  & $1.56\pm0.05$ & $1.22\pm0.02$ \\
16 Cyg B & $5750\pm 50$  & $0.05\pm0.06$  & $1.27\pm0.04$ & $1.12\pm0.02$ \\
\hline
\end{tabular} 
\end{table*}

My first objective in this work is to evaluate the results published in previous studies that already show a good level of agreement \citep{Metcalfe12,Metcalfe15,Bellinger16,Creevey17}. These works adopt various estimation strategies to obtain the stellar parameters. \citet{Metcalfe15} and \citet{Creevey17} consider the problem from a frequentist maximum-likelihood perspective, optimising their criteria using a genetic algorithm \citep{Metcalfe09}. \citet{Bellinger16} opts for a Bayesian approach using neural-network techniques to sample the probability densities of the stellar parameters. In this study, I also adopt the Bayesian approach and use Markov chain Monte Carlo (MCMC) sampling methods to approximate the probability density of the stellar parameters because  the observations are fixed. This method has not been considered in previous studies.

The second objective is to explore the impact of some non-standard processes, namely microscopic diffusion and overshoot. The former is routinely used in solar models because it helps to reproduce properly the sound speed as measured by helioseismology \citep{Basu16}. In this work, the main concern is to compare the effect of including or not the diffusion of metals below the convective envelope.

The method itself and the algorithmic set-ups are described in Section~\ref{sect:method}. In Section~\ref{sect:blabla}, I present the results of the estimation problem and compare these with those of previous studies.

\section{Method}\label{sect:method}

\subsection{Bayesian statistical model and sampling}\label{sect:BSM}

This study was carried out within the Bayesian framework. It is particularly well suited to parameter estimation in the context of stellar physics. Indeed, contrary to laboratory experiments, for operational reasons it is usually not possible to repeat measurements of observable stellar quantities. The classical approach of frequentist statistics postulates that the data are random variables. These are described by probabilities distributions, which often depend on parameters. In these statistical models, any parameter is considered a deterministic quantity. Owing to the lack of repeated measurements, it is nevertheless difficult to estimate these parameters.

Bayesian statistics, thanks to the introduction of the prior distribution of the parameters, reverse this picture. The data become the fixed quantities and the parameters the random quantities. A significant practical advantage is that we can sample the space of parameters using various numerical strategies and, subsequently, use statistical tools to estimate the stellar parameters. This replaces advantageously the need for multiple measurements in classical frequentist statistics.

Bayesian statistics have been described in many monographs and review articles \citep[see e.g.][]{Berger85,Robert05,vonToussaint11}, including in the context of stellar parameter estimation \citep[see e.g.][and references therein]{Bazot08,Bazot12,Bazot16}. In this Section, I simply recall the most important points of the approach and briefly describe the algorithmic strategy chosen for parameter estimation.\\

Central to Bayesian statistics is the Bayes formula
\begin{equation}\label{eq:bayes}
  \displaystyle
  p(\thetav|\xv) \propto p(\xv|\thetav)p(\thetav).
\end{equation}
In this formula $p(\thetav|\xv)$ is called the posterior density of $\thetav$ conditional on $\xv$. This definition is perfectly general, but in the following I identify $\thetav$ to the parameters of the model and $\xv$ to the data. The likelihood is the quantity $f(\thetav) = p(\xv|\thetav)$. Importantly, this is a probability density for the data (i.e. when $p(\xv|\thetav)$ is seen as a function of $\xv$) but not for the parameters. It can be shown in frequentist statistics that a random experiment can be fully specified by the determination of a likelihood function \citep{Birnbaum62,Robert05}. Finally, $p(\thetav)$ is the prior density on the parameters. This is the central concept of Bayesian statistics.  The addition of a prior density is the key element that allows for shifting from the data to parameters as the random quantities. The prior density encodes the information one has on the parameters, $\thetav$, before their inference based on the measurements $\xv$.

Once the priors have been specified, we can make statistical statements based on the posterior density. The estimation of moments of $p(\thetav|\xv)$ and credible intervals requires the integration of the density either over subsets of its domain. Therefore, I wish to obtain an approximation of $p(\thetav|\xv)$, which does not have, in general, a closed form. In order to obtain it, I used an MCMC algorithm similar to that described in \citep{Bazot19}. The adaptive MCMC algorithm \citep{Haario01} was run independently on ten chains. Convergence of the MCMC simulations are discussed in Appendix~\ref{app:conv}.

\subsection{Data and likelihood}\label{sect:likelihood}

In order to set up a Bayesian statistical model, we need to define the likelihood. I first assume that the measurement errors are random and additive. This allows us to write
\begin{equation}\label{eq:statmod}
  \displaystyle
  \xv = \Sv(\thetastv) + \epsilonv,
\end{equation}
where $\Sv$ is a theoretical model, which depends on the parameters $\thetastv$; and $\epsilonv$ is a random vector with zero mean, that is unbiased measurements are assumed, and the variance  is determined using the observations.{Or, depending on your meaning, "...mean, that is unbiased measurements
are assumed and the variance  is determined using the observations.}

For the sake of comparison, the non-seismic measurements (effective temperature, surface luminosity, surface metallicity and radius) I used are the same as those adopted in \citet{Metcalfe15}. Non-seismic constraints are listed in Table~\ref{tab:obs}. Non-seismic measurements are assumed independent, therefore their likelihood is simply the product of the individual likelihoods. These are considered Gaussian, $\mathcal{N}(\mu_k,\sigma^2_k)$, with $\mu_k$ the observed value of the observable $k$ (for a given ordering of the non-seismic observations) and $\sigma^2_k$ its variance.

I chose not to use individual frequencies as a seismic diagnostic because of the need to estimate the surface effects that affect these frequencies \citep{Kjeldsen08}, which are not included in the physical model $\Sv(\thetastv)$. Taking these effects into account can be problematic \citep{Bazot13} and I instead adopt the frequency ratios, $\rzo$, $\rzt$, and $\rot$ defined by \citet{RV03}. These can be straightforwardly computed from Table~S2 of \citet{Davies15}. They are well understood theoretically and it has been emphasised, using empirical arguments, that they indeed seem largely independent of the surface effects \citep{SA13}. I first assume that seismic and non-seismic data are uncorrelated. This means that the likelihood can be written as $f(\thetav) = f(\thetav)_{\mathrm{ns}}f(\thetav)_{\mathrm{s}}$; in this equation, and in the following, ns and s stand for {}non-seismic{} and {}seismic{}, respectively.

The assumption is made that the statistical noise on the measured frequencies is Gaussian and that their covariance matrix is diagonal. This is not the case for the individual ratios, which are in general correlated, since any given pair of separation ratios may involve one or several common eigenfrequencies. I assume that the seismic likelihood is a Gaussian random vector distributed as $\mathcal{N}(0,\Sigmav)$. The covariance matrix $\Sigmav$ has to be evaluated from the variances of the individual frequencies. In this work, I use linear approximations as suggested by \citet{Roxburgh17}. For given orderings of the individual frequency and the frequency ratios $\rzo$, $\rzt$, and $\rot$, the coefficient $c_{i,j}$ of the covariance matrix is given by $c_{i,j} = \partial_mx_{\mathrm{s},i}c_{f;m,n}\partial_nx_{\mathrm{s},j}$, where $c_{f;m,n}$ is a coefficient of the covariance matrix, $\Sigmav_f$, of the frequencies (for 16 Cyg A and B, $c_{f;m,n} = 0$ if $m\neq n$), and the symbol $\partial_m$ indicating derivation with respect to the $m$-th frequency. I note that this covariance matrix could also be approximated using simulated eigenfrequencies sampled from $\mathcal{N}(0,\Sigmav_f)$, simply computing the ratios for each realisation and then evaluating the covariance matrix $\Sigmav$ from the corresponding frequency-ratio sample. In general this method give results in fair agreement with the linear approximation, provided the number of realisations is large enough. I adopt the former method for the sake of simplicity and efficiency. 

Given these considerations, the likelihood function in Eq.~(\ref{eq:bayes}) can be written as
\begin{equation}\label{eq:likelihood}
  \begin{split}
    \displaystyle
    p(\xv|\thetav) \propto & \exp{\left(-\frac{1}{2}\sum_k\frac{(\mathcal{S}_{\mathrm{ns},k}(\thetastv) - x_{\mathrm{ns},k})^2}{\sigma_k^2}\right)}\\
     &\times\exp{\left(-\frac{1}{2}(\xv_\mathrm{s} - \mathcal{S}_{\mathrm{s}}(\thetastv))^T\Sigmav^{-1}(\xv_s - \mathcal{S}_{\mathrm{s}}(\thetastv) )\right)}.
    \end{split}
\end{equation}

The likelihood function departs from that considered in \citet{Metcalfe15} in that the authors did not include the non-diagonal terms in $\Sigmav$ (Metcalfe, private communication). This likelihood function also differs from \citet{Creevey17} since the authors only considered non-diagonal terms in $\Sigmav$ for pairs of $\rzo$ individual ratios, but not for pairs of $\rzt$ ratios or $\rzo$/$\rzt$ pairs. Furthermore, \citet{Metcalfe15} use $\rzoz$ ratios instead of $\rzo$ ratios, the former being obtained by the addition of the $r_{10}$ ratios \citep{RV03} to the latter. This procedure has been criticised by \citet{Roxburgh18} on the grounds that in an $\rzoz$ ratio $2N$ quantities are fitted, while stellar models provide only $N$ frequency phase corresponding to both the $\rzo$ and $r_{10}$ sequences. This may lead to an over-fitting configuration, possibly introducing biases into the estimation process.

\subsection{Physical models and priors}\label{sect:prior}

In Eq.~(\ref{eq:bayes}), the parameters may encapsulate any relevant quantity of the Bayesian statistical model that ought to be estimated. In practice, all the parameters that do not enter as an argument of $\Sv$ are fixed. This includes the $\mu_k$s, the $\sigma_k$, and the coefficients of $\Sigmav$, and therefore $\thetav = \thetastv$.

The physical model I adopt is a spherically symmetric, non-rotating, non-magnetic star. The corresponding equations for the stellar structure and evolution are solved numerically via ASTEC \citep{JCD08a} and those for stellar pulsations via {\tt adipls} \citep{JCD08b}. The equation of state was set according to the OPAL prescription \citep{OPAL02}. The opacities are also provided by the OPAL collaboration \citep{Iglesias96}. Convection is treated following the mixing-length formalism of \citet{BV58}. Nuclear reaction rates are from the NACRE collaboration \citep{Angulo99} and supplemented by the values given in \citet{Angulo05} for the $^{14}$N(p,$\gamma$)$^{15}$O reaction. 

\begin{figure*}[h!]
   \centering
   \includegraphics[width=.45\textwidth]{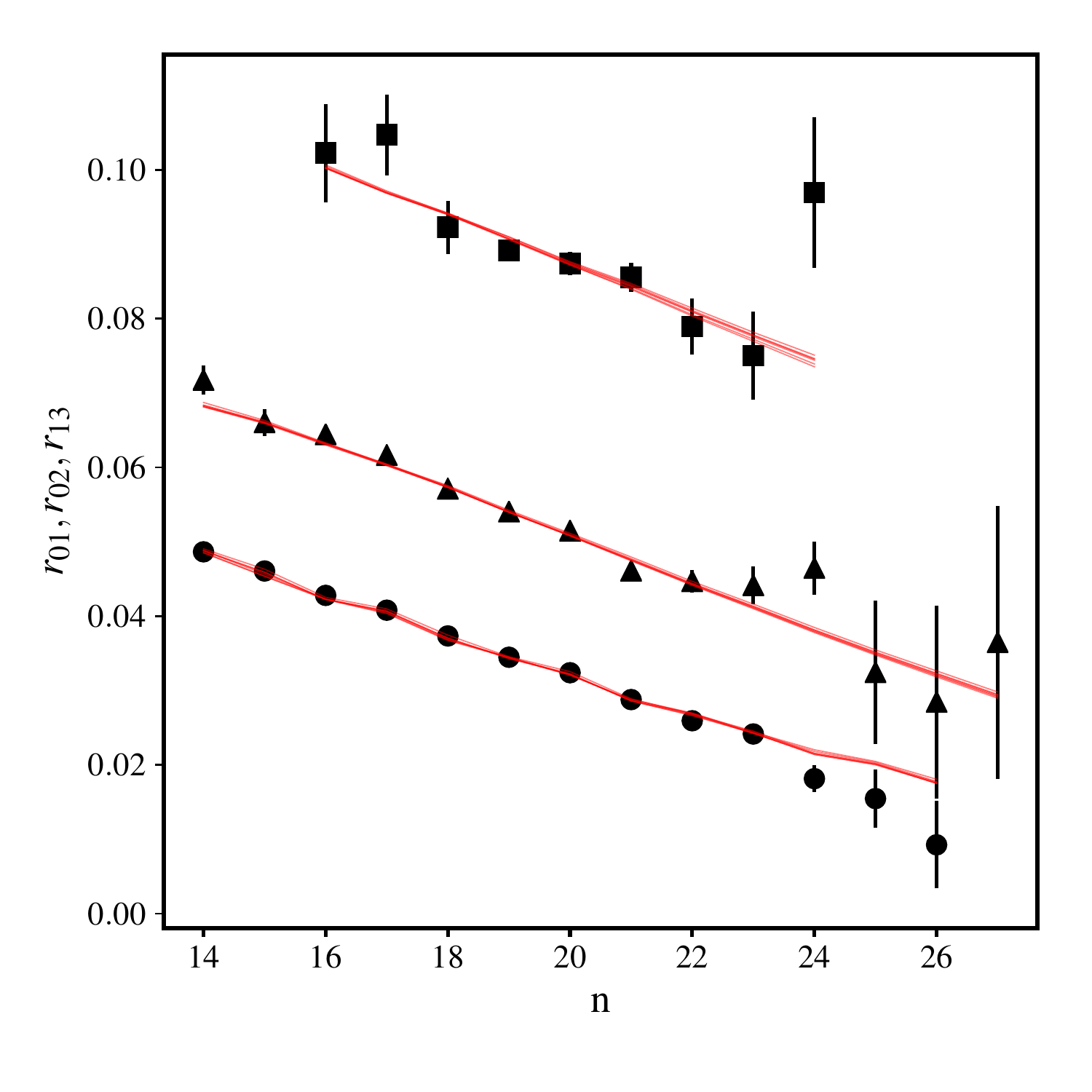}
   \includegraphics[width=.45\textwidth]{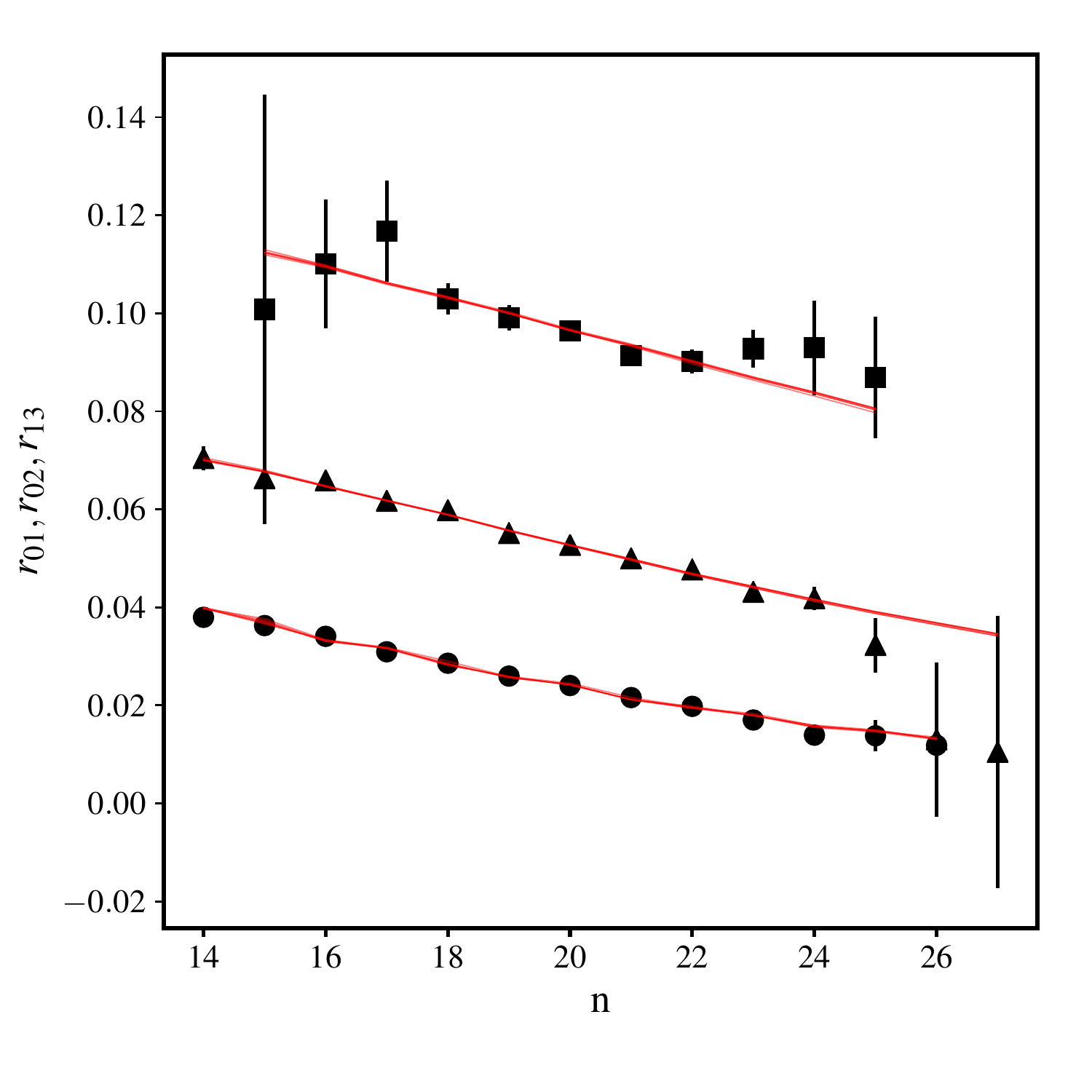}
   \caption{Seismic diagnostic for 16 Cyg A (left panel) and B (right panel). The squares, triangles, and circles denote the observed $r_{13}$, $r_{02}$, and $r_{01}$ observed ratios, respectively. The red lines show the best models for all runs in Tables~\ref{tab:params_16cygA} and \ref{tab:params_16cygB}.}
         \label{fig:seismic_observations}
\end{figure*}

Other physical prescriptions were varied to test their impact on parameter estimation. Microscopic diffusion of elements heavier than hydrogen is treated using the formalism of \citet{Michaud93}. I either considered simple helium diffusion or helium and metal diffusion. The ratios of chemical element abundances are taken either from \citet{Grevesse93} or \citet{Grevesse98}. For the sake of simplicity and comparison with \citet{Metcalfe15} and \citet{Creevey17}, I disregarded the abundance ratios provided by \citet{Asplund09}, which have raised many difficulties in the solar case \citep[see e.g.][]{Guzik06,Castro07,Basu08,Antia11,Gough12,Basu16}. This case is left to future studies. Finally, I considered the possibility of overshoot below the convective envelope, that is transport process beyond the limit point predicted by a linear stability analysis.

Solving the equations for stellar structure and evolution under the above assumptions demand to set five free parameters, plus one in case overshoot is included. The basic parameters are the stellar mass, $\stmass$, age, $\stage$, initial chemical composition, given by the initial hydrogen-mass fraction, $X_0$, and metallicity, $Z_0$, and the mixing-length parameter, $\alpha$. This latter is a proportionality coefficient between the mean-free path of a fluid parcel in the convective zone and the pressure scale height. The mixing-length parameter sets the depth of the convective zone.
When it is included, a new parameter $\alphaov$ must also be taken into account, whose signification is discussed in Sect.~\ref{sect:overshoot}.

The last necessary step in the construction of the Bayesian statistical model is to specify the priors on the stellar parameters. Following \citet{Bazot18} I assume uniform priors on all parameters. Their boundaries are given in Table~\ref{tab:priors}

\begin{table}
\begin{center}
\caption{Lower and upper bounds used for the prior uniform densities for each stellar parameter.}
\label{tab:priors}
\begin{tabular}{@{}lcc@{}}
\hline
\hline
Parameter&  Lower bound & Upper bound\\
\hline
$M$ (M$_{\odot}$)      & 0.7    & 1.25\\
$\stage$ (Gyr)        &  0.001 & 13\\
$Z_0$                 & 0.010  & 0.035\\
$X_0$                 & 0.525  & 0.750\\
$\alpha$              & 1.0    & 3.5 \\
$\alphaov$            & 0.0    & 0.3 \\
\hline
\end{tabular}
\end{center}
\end{table}

\section{Results and discussion}\label{sect:blabla}

\begin{sidewaystable*}
\caption{Set-up of the MCMC simulations and physical quantities inferred using the ASTEC stellar evolution code for 16~Cyg~A. The first column gives the seismic indicators retained as constraints. The second column gives the chosen  abundance mixture with GN93 for \citet{Grevesse93} and GS98 for \citet{Grevesse98}. The third column indicates the elements included in the diffusion equation. The fourth column specifies, if included, the nature of overshoot. Columns 5 -- 10 show the estimates of the stellar parameters. The last column gives the estimates for the initial helium-mass fraction. For each parameter the first line gives the estimate in the sense of the maximum of the marginal density, with credible intervals. The second line reports the MAP estimates. The third line gives the PM and PSD estimates.}

  \label{tab:params_16cygA}      
\centering                                      
\begin{tabular}{lccccccccccc}          
\hline\hline
\\[-2.mm]
\# & Seismic indicators & Abundances & Diffusion & Overshoot  & $M/M_{\odot}$ & $\stage$ (Gyr) & $X_0$ & $Z_0$ & $\alpha$ & $\alpha_{\mathrm{ov}}$ & $Y_0$\\[1mm]
\hline
\\[-2.mm]
& & & & & $1.07^{+0.02}_{-0.02}$ & $6.91^{+0.21}_{-0.19}$ & $0.718^{+0.012}_{-0.012}$ & $0.0217^{+0.0018}_{-0.0015}$ & $2.16^{+0.07}_{-0.09}$ & -- & ${0.260}^{+0.012}_{-0.011}$\\[1.5mm]
\multirow{1}{*}{1} & \multirow{1}{*}{$\rzo$, $\rzt$, $\rot$}  & GN93 & Y & No &$1.07$ & $6.88$ & $0.718$ & $0.0219$ & $2.19$ & -- & $0.260$\\[1mm]
& &&&&  1.07 (0.02) & 6.93 (0.19) & 0.719 (0.012) & 0.0220 (0.0016) & 2.16 (0.07) & -- & 0.259 (0.011)\\
\hline
\\[-2.mm]
& & & & & $1.07^{+0.02}_{-0.02}$ & $6.77^{+0.22}_{-0.17}$ & $0.707^{+0.013}_{-0.012}$ & $0.0218^{+0.0018}_{-0.0014}$ & $2.14^{+0.09}_{-0.06}$ & -- & ${0.271}^{+0.011}_{-0.013}$\\[1.5mm]
\multirow{1}{*}{2} & \multirow{1}{*}{$\rzo$, $\rzt$, $\rot$}  & GS98 & Y & No &$1.06$ & $6.78$ & $0.705$ & $0.0213$ & $2.17$ & -- & $0.274$\\[1mm]
& &&&&  1.07 (0.02) & 6.80 (0.19) & 0.708 (0.012) & 0.0220 (0.0015) & 2.16 (0.07) & -- & 0.270 (0.011)\\
\hline
\\[-2.mm]
& & & & & $1.08^{+0.02}_{-0.02}$ & $7.25^{+0.19}_{-0.20}$ & $0.717^{+0.011}_{-0.013}$ & $0.0241^{+0.0019}_{-0.0016}$ & $2.04^{+0.07}_{-0.07}$ & -- & ${0.259}^{+0.012}_{-0.011}$\\[1.5mm]
\multirow{1}{*}{3} & \multirow{1}{*}{$\rzo$, $\rzt$, $\rot$}  & GN93 & No & No &$1.08$ & $7.27$ & $0.716$ & $0.0247$ & $2.03$ & -- & $0.260$\\[1mm]
& &&&&  1.08 (0.02) & 7.25 (0.19) & 0.716 (0.011) & 0.0244 (0.0017) & 2.04 (0.07) & -- & 0.260 (0.011)\\
\hline
\\[-2.mm]
& & & & & $1.07^{+0.02}_{-0.02}$ & $6.79^{+0.19}_{-0.20}$ & $0.708^{+0.013}_{-0.011}$ & $0.0241^{+0.0018}_{-0.0019}$ & $2.12^{+0.08}_{-0.08}$ & -- & ${0.266}^{+0.012}_{-0.011}$\\[1.5mm]
\multirow{1}{*}{4} & \multirow{1}{*}{$\rzo$, $\rzt$, $\rot$}  & GN93 & Y + Z & No &$1.07$ & $6.78$ & $0.708$ & $0.0238$ & $2.14$ & -- & $0.268$\\[1mm]
& &&&&  1.07 (0.02) & 6.80 (0.19) & 0.709 (0.012) & 0.0242 (0.0018) & 2.12 (0.07) & -- & 0.266 (0.011)\\
\hline
\\[-2.mm]
& & & & & $1.07^{+0.02}_{-0.02}$ & $6.91^{+0.20}_{-0.17}$ & $0.718^{+0.013}_{-0.010}$ & $0.0217^{+0.0017}_{-0.0014}$ & $2.15^{+0.07}_{-0.07}$ & $0.020^{+0.029}_{-0.017}$ & ${0.258}^{+0.011}_{-0.010}$\\[1.5mm]
\multirow{1}{*}{5} & \multirow{1}{*}{$\rzo$, $\rzt$, $\rot$} & GN93 & Y & Penetrative &$1.07$ & $6.84$ & $0.718$ & $0.0216$ & $2.19$ & $0.012$ & $0.260$\\[1mm]
& &&&&  1.07 (0.02) & 6.93 (0.17) & 0.720 (0.011) & 0.0219 (0.0015) & 2.15 (0.07) & 0.035 (0.024) & 0.258 (0.010)\\
\hline
\\[-2.mm]
& & & & & $1.07^{+0.02}_{-0.02}$ & $6.78^{+0.19}_{-0.19}$ & $0.710^{+0.012}_{-0.013}$ & $0.0239^{+0.0019}_{-0.0017}$ & $2.12^{+0.08}_{-0.08}$ & $0.030^{+0.045}_{-0.025}$ & ${0.265}^{+0.012}_{-0.011}$\\[1.5mm]
\multirow{1}{*}{6}& \multirow{1}{*}{$\rzo$, $\rzt$, $\rot$} & GN93 & Y + Z & Penetrative &$1.07$ & $6.86$ & $0.712$ & $0.0237$ & $2.09$ & $0.001$ & $0.265$\\[1mm]
& &&&&  1.07 (0.02) & 6.79 (0.18) & 0.710 (0.012) & 0.0242 (0.0017) & 2.12 (0.07) & 0.052 (0.035) & 0.266 (0.011)\\
\hline
\\[-2.mm]
& & & & & $1.04^{+0.02}_{-0.02}$ & $6.98^{+0.10}_{-0.08}$ & $0.704^{+0.008}_{-0.007}$ & $0.0215^{+0.0012}_{-0.0011}$ & $2.08^{+0.06}_{-0.05}$ & -- & ${0.275}^{+0.007}_{-0.009}$\\[1.5mm]
\multirow{1}{*}{7} & \multirow{1}{*}{$\rzoz$, $\rzt$, $\rot$}  & GN93 & Y & No &$1.03$ & $7.01$ & $0.702$ & $0.0220$ & $2.07$ & -- & $0.276$\\[1mm]
& &&&&  1.04 (0.02) & 6.99 (0.09) & 0.704 (0.007) & 0.0217 (0.0011) & 2.08 (0.05) & -- & 0.274 (0.008)\\
\hline
\end{tabular}
\\[3.mm]
\end{sidewaystable*}

\begin{sidewaystable*}
\caption{Same as Table~\ref{tab:params_16cygA} for 16 Cyg B.}

  \label{tab:params_16cygB}      
\centering                                      
\begin{tabular}{lccccccccccc}          
\hline\hline
\\[-2.mm]
\# & Seismic indicators & Abundances & Diffusion & Overshoot  & $M/M_{\odot}$ & $\stage$ (Gyr) & $X_0$ & $Z_0$ & $\alpha$ & $\alpha_{\mathrm{ov}}$ & $Y_0$\\[1mm]
\hline
\\[-2.mm]
& & & & & $1.05^{+0.02}_{-0.02}$ & $6.81^{+0.20}_{-0.17}$ & $0.730^{+0.010}_{-0.015}$ & $0.0217^{+0.0021}_{-0.0016}$ & $2.16^{+0.07}_{-0.08}$ & -- & ${0.247}^{+0.014}_{-0.009}$\\[1.5mm]
\multirow{1}{*}{1}& \multirow{1}{*}{$\rzo$, $\rzt$, $\rot$}  & GN93 & Y & No &$1.06$ & $6.99$ & $0.740$ & $0.0204$ & $2.13$ & -- & $0.240$\\[1mm]
& &&&&  1.05 (0.02) & 6.83 (0.17) & 0.727 (0.012) & 0.0221 (0.0018) & 2.16 (0.07) & -- & 0.251 (0.011)\\
\hline
\\[-2.mm]
& & & & & $1.05^{+0.02}_{-0.02}$ & $6.71^{+0.21}_{-0.18}$ & $0.721^{+0.010}_{-0.017}$ & $0.0216^{+0.0022}_{-0.0015}$ & $2.14^{+0.09}_{-0.06}$ & -- & ${0.258}^{+0.015}_{-0.011}$\\[1.5mm]
\multirow{1}{*}{2}& \multirow{1}{*}{$\rzo$, $\rzt$, $\rot$}  & GS98 & Y & No &$1.04$ & $6.77$ & $0.720$ & $0.0207$ & $2.15$ & -- & $0.259$\\[1mm]
& &&&&  1.05 (0.02) & 6.72 (0.18) & 0.718 (0.013) & 0.0221 (0.0018) & 2.16 (0.07) & -- & 0.260 (0.012)\\
\hline
\\[-2.mm]
& & & & & $1.06^{+0.02}_{-0.02}$ & $7.17^{+0.19}_{-0.20}$ & $0.723^{+0.012}_{-0.014}$ & $0.0238^{+0.0021}_{-0.0018}$ & $2.03^{+0.07}_{-0.07}$ & -- & ${0.252}^{+0.013}_{-0.011}$\\[1.5mm]
\multirow{1}{*}{3}& \multirow{1}{*}{$\rzo$, $\rzt$, $\rot$}  & GN93 & No & No &$1.06$ & $7.21$ & $0.725$ & $0.0234$ & $2.03$ & -- & $0.251$\\[1mm]
& &&&&  1.06 (0.02) & 7.16 (0.18) & 0.722 (0.012) & 0.0241 (0.0019) & 2.04 (0.06) & -- & 0.254 (0.012)\\
\hline
\\[-2.mm]
& & & & & $1.05^{+0.02}_{-0.02}$ & $6.65^{+0.18}_{-0.18}$ & $0.709^{+0.016}_{-0.010}$ & $0.0249^{+0.0021}_{-0.0020}$ & $2.13^{+0.06}_{-0.08}$ & -- & ${0.265}^{+0.010}_{-0.015}$\\[1.5mm]
\multirow{1}{*}{4}& \multirow{1}{*}{$\rzo$, $\rzt$, $\rot$}  & GN93 & Y + Z & No &$1.05$ & $6.77$ & $0.721$ & $0.0235$ & $2.10$ & -- & $0.256$\\[1mm]
& &&&&  1.05 (0.02) & 6.65 (0.17) & 0.712 (0.013) & 0.0250 (0.0019) & 2.12 (0.07) & -- & 0.263 (0.012)\\
\hline
\\[-2.mm]
& & & & & $1.05^{+0.02}_{-0.02}$ & $6.82^{+0.20}_{-0.17}$ & $0.728^{+0.012}_{-0.013}$ & $0.0218^{+0.0019}_{-0.0016}$ & $2.16^{+0.07}_{-0.08}$ & $0.010^{+0.022}_{-0.008}$ & ${0.250}^{+0.011}_{-0.012}$\\[1.5mm]
\multirow{1}{*}{5}& \multirow{1}{*}{$\rzo$, $\rzt$, $\rot$} & GN93 & Y & Penetrative &$1.05$ & $6.84$ & $0.731$ & $0.0211$ & $2.16$ & $0.001$ & $0.248$\\[1mm]
& &&&&  1.05 (0.02) & 6.83 (0.17) & 0.727 (0.011) & 0.0220 (0.0017) & 2.16 (0.07) & 0.022 (0.015) & 0.251 (0.011)\\
\hline
\\[-2.mm]
& & & & & $1.05^{+0.02}_{-0.02}$ & $6.65^{+0.19}_{-0.18}$ & $0.713^{+0.013}_{-0.014}$ & $0.0248^{+0.0021}_{-0.0020}$ & $2.12^{+0.07}_{-0.08}$ & $0.014^{+0.023}_{-0.011}$ & ${0.262}^{+0.013}_{-0.012}$\\[1.5mm]
\multirow{1}{*}{6} & \multirow{1}{*}{$\rzo$, $\rzt$, $\rot$} & GN93& Y + Z & Penetrative&$1.05$ & $6.67$ & $0.711$ & $0.0243$ & $2.12$ & $0.003$ & $0.265$\\[1mm]
& &&&&  1.05 (0.02) & 6.66 (0.17) & 0.713 (0.013) & 0.0250 (0.0019) & 2.12 (0.07) & 0.025 (0.017) & 0.262 (0.012)\\
\hline
\\[-2.mm]
& & & & & $1.02^{+0.02}_{-0.02}$ & $6.91^{+0.10}_{-0.08}$ & $0.712^{+0.008}_{-0.007}$ & $0.0220^{+0.0012}_{-0.0013}$ & $2.07^{+0.05}_{-0.05}$ & -- & ${0.265}^{+0.008}_{-0.008}$\\[1.5mm]
\multirow{1}{*}{7}& \multirow{1}{*}{$\rzoz$, $\rzt$, $\rot$}  & GN93 & Y & No &$1.02$ & $6.88$ & $0.709$ & $0.0217$ & $2.10$ & -- & $0.269$\\[1mm]
& &&&&  1.02 (0.02) & 6.92 (0.09) & 0.712 (0.007) & 0.0220 (0.0012) & 2.07 (0.05) & -- & 0.266 (0.007)\\
\hline
\end{tabular}
\\[3.mm]
\end{sidewaystable*}

The results of the MCMC simulations are given in Tables~\ref{tab:params_16cygA} and \ref{tab:params_16cygB} for 16~Cyg~A and B, respectively. For both stars I used $\rzo$, $\rzt$, and $\rot$ as seismic diagnostics for the first five runs. For the sixth run, $\rzo$ was replaced with $\rzoz$ to test the potential effects of over-fitting. For each simulation I provide three different estimates of the stellar parameters parameters $\stmass$, $\stage$, $X_0$, $Z_0$, $\alpha$ and, when relevant, $\alphaov$. I also provide estimates for the initial helium-mass fraction, $Y_0$, to facilitate comparison with other studies, this parameter often being reported instead of $X_0$. For a given set-up, the first line of estimates correspond to the maxima of the marginal densities. These maxima are given alongside credible intervals that are defined as the smallest interval with probability 0.683 that contains the maximum. The second line is the maximum a posteriori (MAP) obtained from the joint  distribution $p(\thetastv|\xv)$. I do not provide uncertainties for this point estimate. In the third  the posterior means (PM) and posterior standard deviations (PSD) are reported; these estimates are computed from the marginal densities.

The baseline case was chosen so that it corresponds to the set-up used in \citet{Metcalfe15}, who also used ASTEC\footnote{The full results of the runs can be found at \url{https://amp.phys.au.dk/browse/simulation/767} for 16 Cyg A and \url{https://amp.phys.au.dk/browse/simulation/768} for 16 Cyg B.}. This set-up uses the \citet{Grevesse93} abundances ratios. Only helium is included in the diffusion equation of chemical elements. Overshoot is not taken into account.

\begin{figure*}[h!]
   \centering
   \includegraphics[width=.45\textwidth]{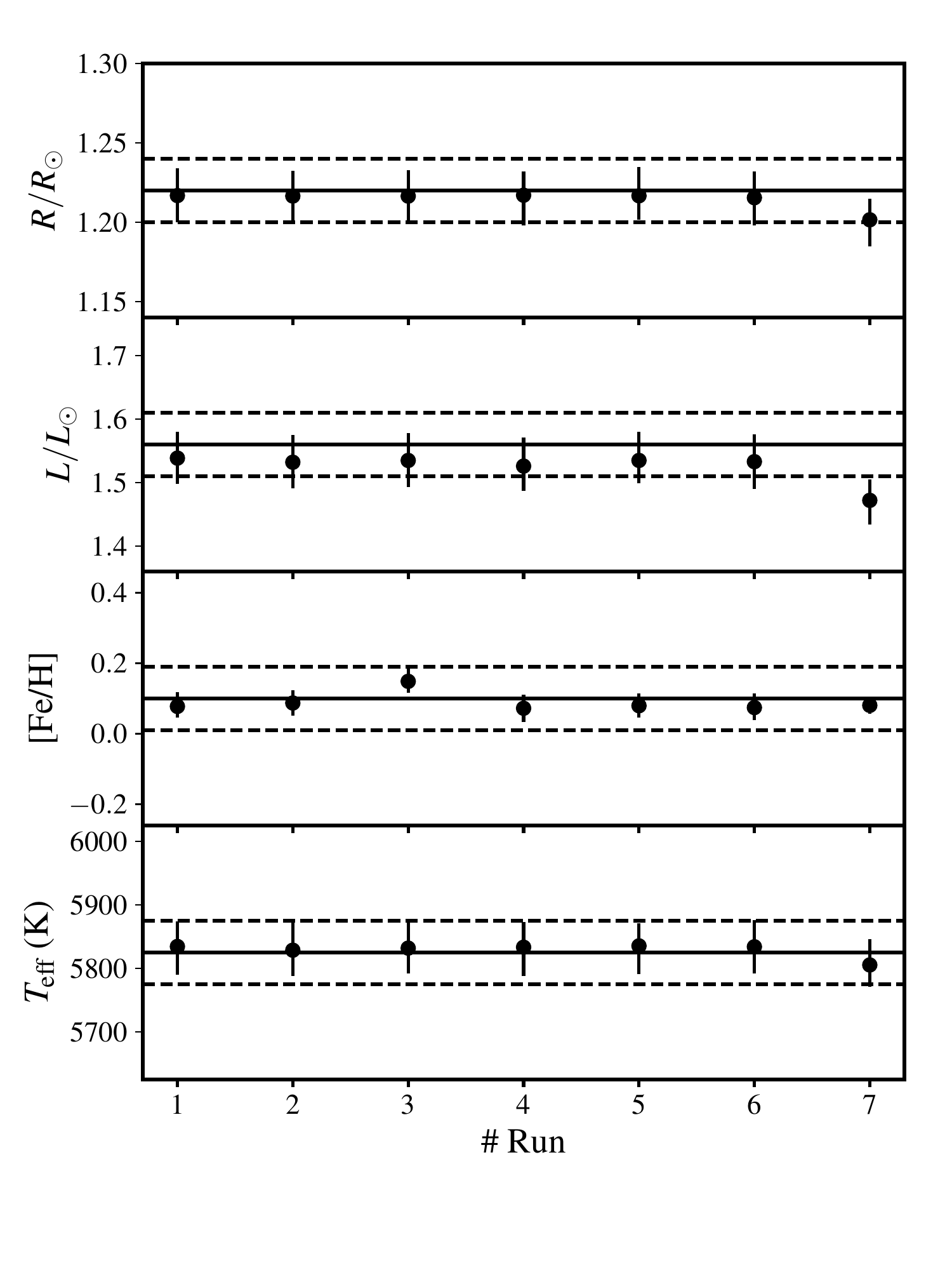}
   \includegraphics[width=.45\textwidth]{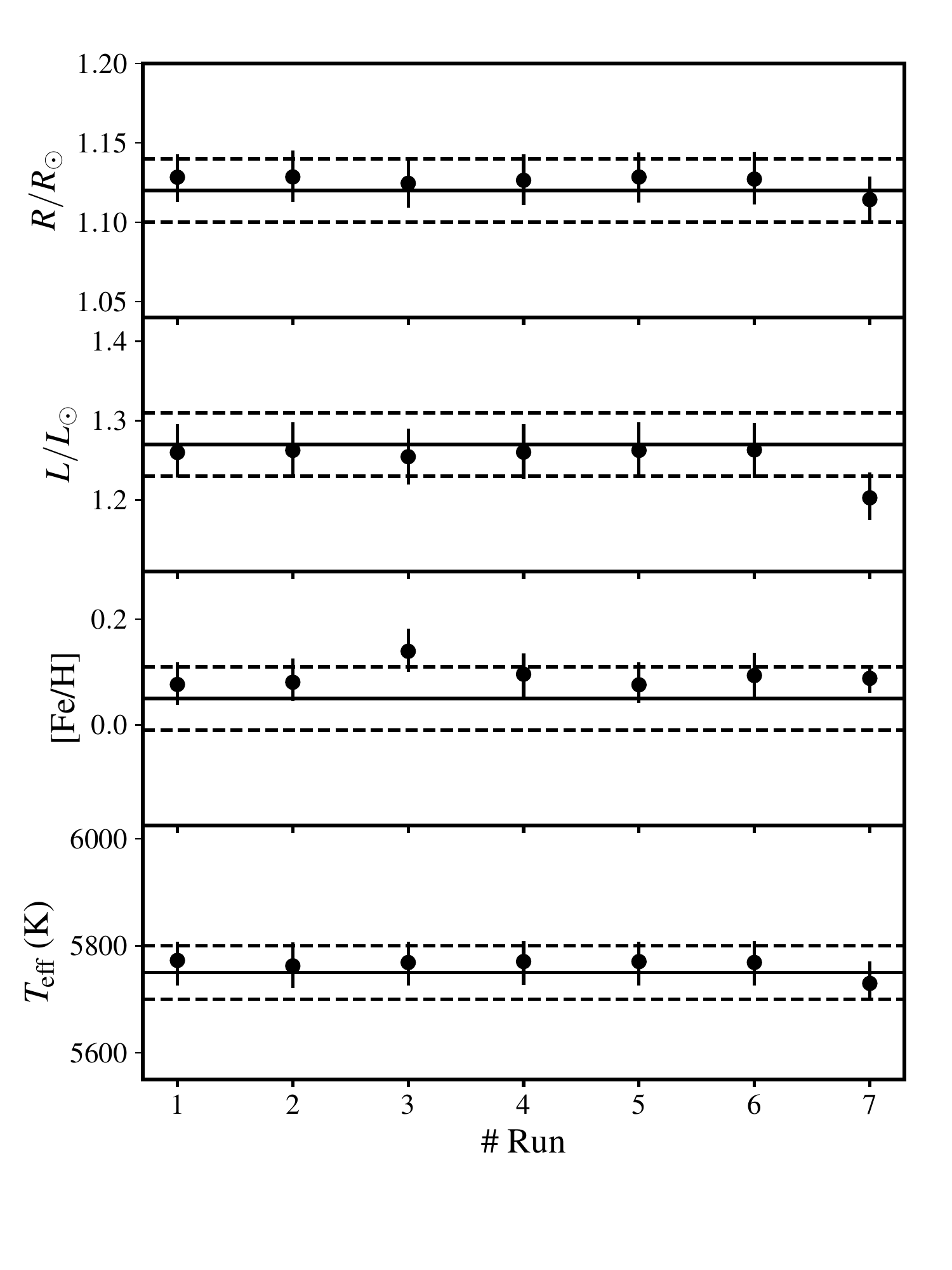}
   \caption{Non-seismic observations for 16 Cyg A (left panel) and B (right panel). The horizontal lines represent the observed values (full lines) and the corresponding uncertainties (dashed lines). The dots denote the estimated values (MAPs of the marginal densities) for the corresponding observation, for each run. The runs are labelled on the abscissa according to the numbering given in Tables~\ref{tab:params_16cygA} and \ref{tab:params_16cygB}.
   }
         \label{fig:ns_observations}
\end{figure*}
The observed seismic diagnostics are indicated in Fig.~\ref{fig:seismic_observations}. I also show the best model for each case in Tables~\ref{tab:params_16cygA} and \ref{tab:params_16cygB} (red thin lines). These optimal theoretical values are barely distinguishable from one model to the other. This indicates that they may be equally valid in order to reproduce the data, and that statistical model comparison methods \cite[see e.g.][Chap.~7]{Robert07} alone will not suffice to distinguish them. Further discussions on model comparison are given in Sects.~\ref{sect:overshoot} and \ref{sect:diffusion}. Visual inspection of Fig.~\ref{fig:seismic_observations} also indicates that the accuracy of the present results is similar to what was obtained by \citet{Metcalfe15}. Figure~\ref{fig:ns_observations} shows the MAP estimates of all non-seismic observations. They are all reproduced with good accuracy, within 1$\sigma$ of the observed values.

\begin{figure*}[h!]
   \centering
   \includegraphics[width=.45\textwidth]{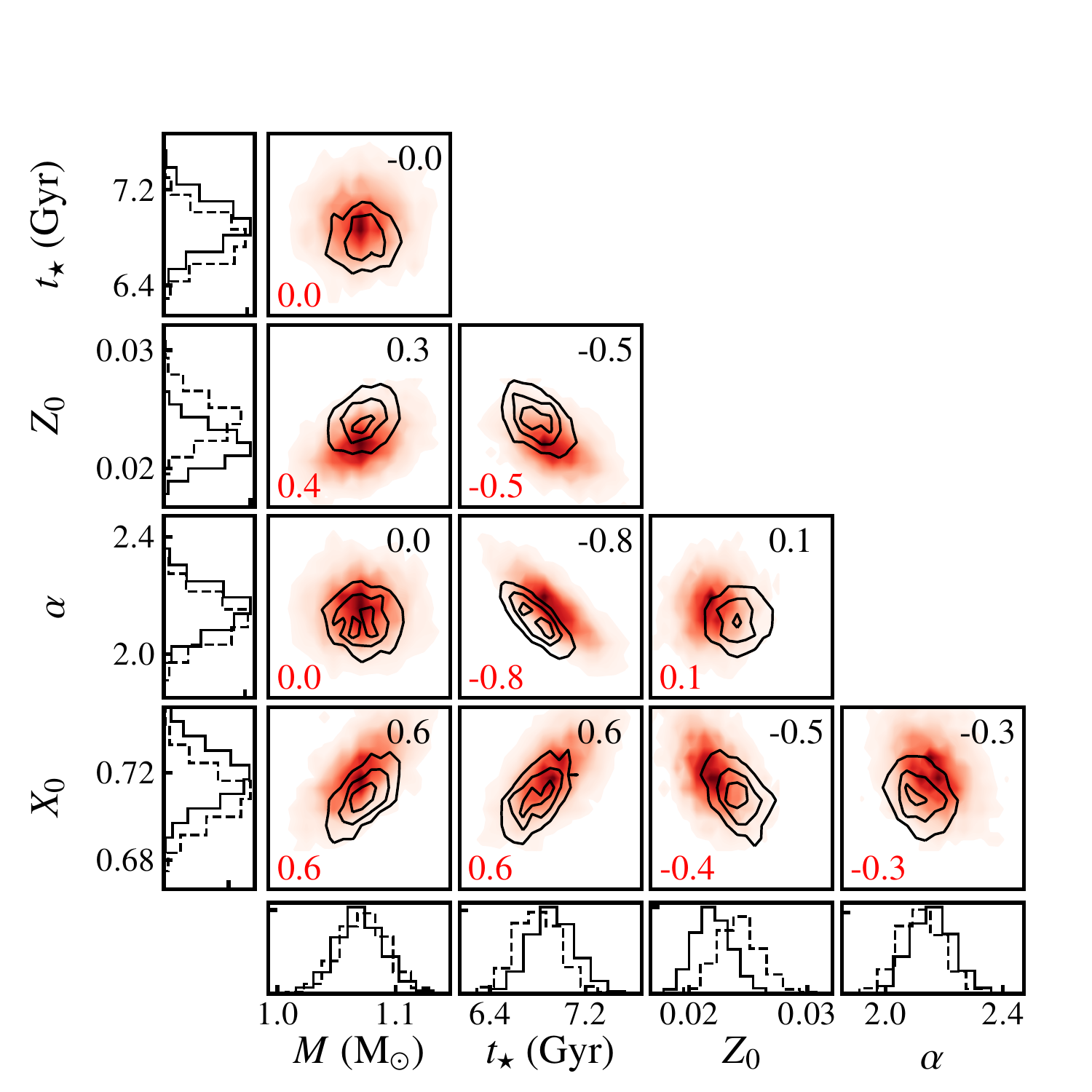}
   \includegraphics[width=.45\textwidth]{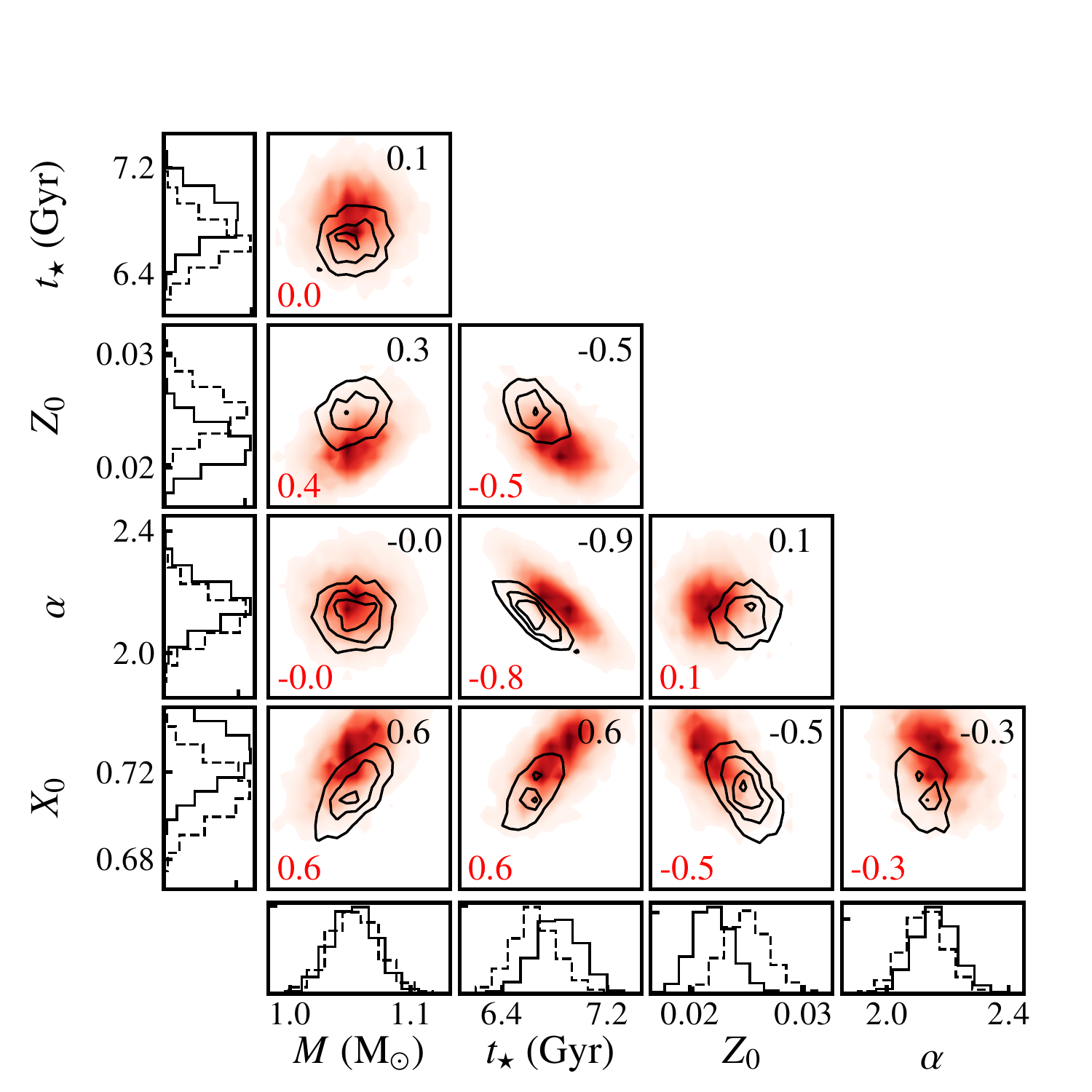}
   \caption{Marginal densities for the stellar parameters $M$, $\stage$, $X_0$, $Z_0$, and $\alpha$ of 16~Cyg~A (left) and B (right). The central panels show the joint marginal densities of the paired parameters. In the side panels are plotted the individual marginal densities. The red shaded areas in the central panels and the full lines in the side panels represent posterior densities for the baseline case, that is without diffusion of metals. The black contours in the central panels and the dashed lines in the side panels represent posterior densities for models with diffusion of metals. The numbers in each panel give the Pearson correlation coefficient for the two variables. The values in red correspond to the red shaded densities and the values in black to the black contours.
   }
         \label{fig:joint-16CygAB-stellar-diffusion}
\end{figure*}

The two-dimensional joint PDFs for this set-up are shown in Fig.~\ref{fig:joint-16CygAB-stellar-diffusion} alongside the corresponding correlation coefficients. The mass correlates most significantly with $X_0$. The age correlates extremely well with the mixing-length parameter and slightly less with $X_0$. It may be counter-intuitive that the mass does not correlate strongly with the age. A likely explanation is that, due to the level of constraint imposed by the seismic data, these two parameters cannot vary enough to allow such a correlation to be observed. Interpreting these correlations is not straightforward because there is in general no one-to-one correspondence between a parameter and an observation. Thus elements of interpretation can be gathered by looking at the correlation between the stellar parameters and other quantities. The mass and  $X_0$ (Pearson correlation
coefficients $\gtrsim 0.8$) both correlate strongly with the luminosity and the radius on the zero-age main sequence (ZAMS). This is related to the mass-luminosity relation on the ZAMS, in which the mean-molecular weight enters \cite[see e.g.][]{Clayton68}. To maintain a roughly constant luminosity, an increase in mass must be compensated by a decrease in the mean-molecular weight. This latter corresponds to an increase in the initial hydrogen-mass fraction and a decrease of the initial helium-mass fraction because the two quantities are perfectly anti-correlated. The interaction between the mass and the hydrogen-mass fraction thus sets the initial conditions of the stellar evolution sequence. Such a behaviour was already noted in \citet{Bazot18}, even though the trend is not as clear in this work.

The mixing-length parameter correlates very strongly with the effective temperature. It is well known that these two quantities are related \cite{Clayton68}. Again, the mixing-length parameter allows for setting the initial conditions of the evolutionary sequence. The age also anti-correlates well with the effective temperature (Pearson correlation coefficient $\sim$0.8), more than with any other observable. This is an evolutionary effect. Those two correlations explain the relation between the age and the mixing length.

  Another very strong correlation of note is that between the initial metallicity and the surface metallicity-to-hydrogen ratio. It is much stronger (Pearson correlation coefficient $> 0.9$) than the correlation between $Z/X$ and $X_0$. This is because too large variations of $X_0$ would also affect the radius and the luminosity, whereas variations of $Z_0$ do not have the same effect. These are very general observations. In order to better understand the behaviour of these correlations with the observational constraints, it would be necessary to use simulated data in the spirit of \citet{Brown94} or \citet{Creevey07}. In Fig.~\ref{fig:joint-16CygAB-stellar-diffusion} also shows one-dimensional marginal densities, which are all Gaussian to a very good approximation. This means that their PSD can be interpreted as a 0.683 credible interval as defined above, the PM and maxima of the marginal being equal.

As is shown below, the stellar mass is of importance as it is a good marker of the robustness of the asteroseismic (and interferometric) constraints. For the baseline case I obtain $1.07\pm0.02$~{\msol} and $1.05\pm0.02$~{\msol} for 16 Cyg A and B. These are fairly close to the estimates of \citet{Metcalfe15} and \citet{Creevey17}, both of which use ASTEC, albeit with slightly different set-up for the latter. The same is true for the other parameters with the possible exception of the age of 16 Cyg A quoted in \citet{Creevey17}, which is higher by 6.5\% (i.e. roughly 2.4$\sigma$ away from the PM value). The high precisions on all parameters can be traced back to the use of asteroseismology. This can be seen qualitatively by comparing the current results to those of \citet{Bazot18} on 18 Sco. This latter star, a solar twin, has even more precise spectrophotometric and interferometric data than 16 Cyg A and B. However, the precision achieved on the estimates of its parameter using much poorer asteroseismic data is far worse. For the age, there is an order of magnitude difference in precision.

A difficulty arises when comparing the uncertainties on these parameters. Indeed both \citet{Metcalfe15} and \citet{Creevey17} use samples obtained from a genetic algorithm to estimate confidence intervals \citep{Metcalfe14}. In the case of MCMC algorithms, asymptotic properties of Markov chains are used to show that the resulting sample is generated according to the target distribution, i.e. $p(\thetastv|\xv)$ in this work. The issue with genetic algorithms is that there exist no such theorem guaranteeing the convergence of the resulting sample to the joint probability of the parameters being optimised\footnote{It is important to be careful with the terminology. Genetic algorithms are used for optimisation. Therefore even considering a convergence towards a probability density from a genetic algorithm is in contradiction with the purpose of the procedure itself.}. Therefore, and even though the quoted uncertainties in these previous studies are often in good numerical agreement with those given in this work, we cannot draw strong conclusions by simply comparing their estimates with those of Tables~\ref{tab:params_16cygA} and \ref{tab:params_16cygB}.

Other studies have provided estimates for the stellar parameters of 16 Cyg A and B. In \citet{SA17} results are reported from multiple groups using various codes and estimation strategies. The values quoted in this study bracket those represented in Tables~\ref{tab:params_16cygA} and \ref{tab:params_16cygB}. Finally, other works to be mentioned are those of \citet{Bellinger16} and \citet{Bellinger17}. In the first paper, the author used neural-network strategies to estimate stellar parameters in a Bayesian fashion. Their estimates are relatively close to those given in this study. In particular, the age estimates are exactly the same for both 16 Cyg A and B. An exception is the mass of 16 Cyg B, which is slightly lower than that seen in Table~\ref{tab:params_16cygB}. As discussed in Sect.~\ref{sect:ratios}, this can potentially be explained by the treatment of the seismic constraints. Their estimated uncertainty on the age of 16 Cyg A is twice as large as those given in Table~\ref{tab:params_16cygB}. Such a discrepancy is not straightforward to explain. A possible factor is the inclusion in their model of a parameter that controls the efficiency of diffusion. However, we would expect 16 Cyg B to be also be affected to some extent, which does not seem to be the case. This problem will require further work. Finally, their distribution for the initial helium-mass fraction of 16 Cyg A shows a bimodality that I do not observe. This stresses that, despite a satisfying general agreement between the two methods, the posterior densities used to derive the stellar parameters still show discrepancies that ought to be explained. 

The estimates of the age for both stars are in extremely good agreement, 6.93$\pm$0.19~Gyr and 6.83$\pm$0.17~Gyr. Confirming previous studies, given the independent modelling of 16 Cyg A and B this
corresponds to an age of 6.88$\pm$0.12~Gyr  for the system. This value is close to that claimed by \citet{Metcalfe15}, roughly departing by $1\sigma$.

\subsection{Chemical abundances and microscopic diffusion}\label{sect:diffusion}

The treatment of chemical abundances may have a strong impact on the physical characteristics of the stars. In this section, I study the potential biases induced by different prescriptions for the abundances ratios and treatments of microscopic diffusion. The results are listed in Tables~\ref{tab:params_16cygA} and \ref{tab:params_16cygB}. The first interesting result is that the mass estimate is remarkably stable with respect to such changes. This reflects the fact that the radius and density, through the seismic data, are well constrained, hence providing a robust mass estimate (see for instance \citeauthor{Bazot12} \citeyear{Bazot12}, \citeauthor{Bazot18} \citeyear{Bazot18} for another example of robust mass estimate using interferometry and asteroseismology; see also \citeauthor{Creevey07} \citeyear{Creevey07} and \citeauthor{Cunha07} \citeyear{Cunha07} for more general discussions on the interplay between asteroseismology and interferometry).

The main impact of the change from \citet{Grevesse93} to \citet{Grevesse98} abundance ratios is a small decrease in $X_0$ and $\stage$, coupled to a small increase of $Y_0$ (all below the 1$\sigma$ uncertainties from the baseline case). The estimate of $Z_0$ remains stable. Overall, the biases to be expected from such a change should remain small. This may not be the case if I had considered \citet{Asplund05} abundances. These are known to produce discrepancies between solar models and helioseismic observations \citep[see][for a review]{Basu16}. Because of these discrepancies I did not include these abundances in this study; the models used in this work were satisfyingly validated in the solar case for the \citet{Grevesse93} and \citet{Grevesse98} abundance ratios.

\begin{figure*}[h!]
   \centering
   \includegraphics[width=.45\textwidth]{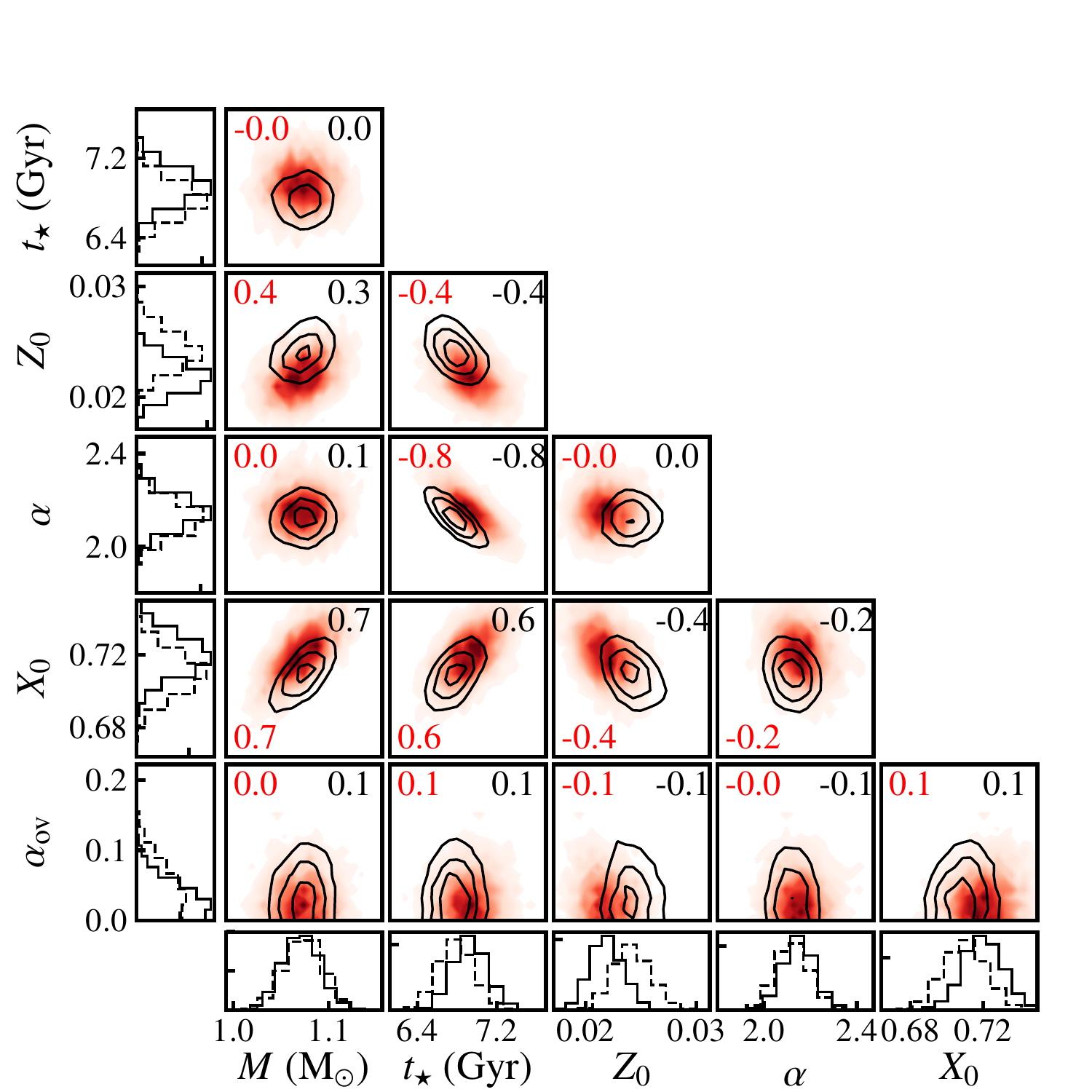}
   \includegraphics[width=.45\textwidth]{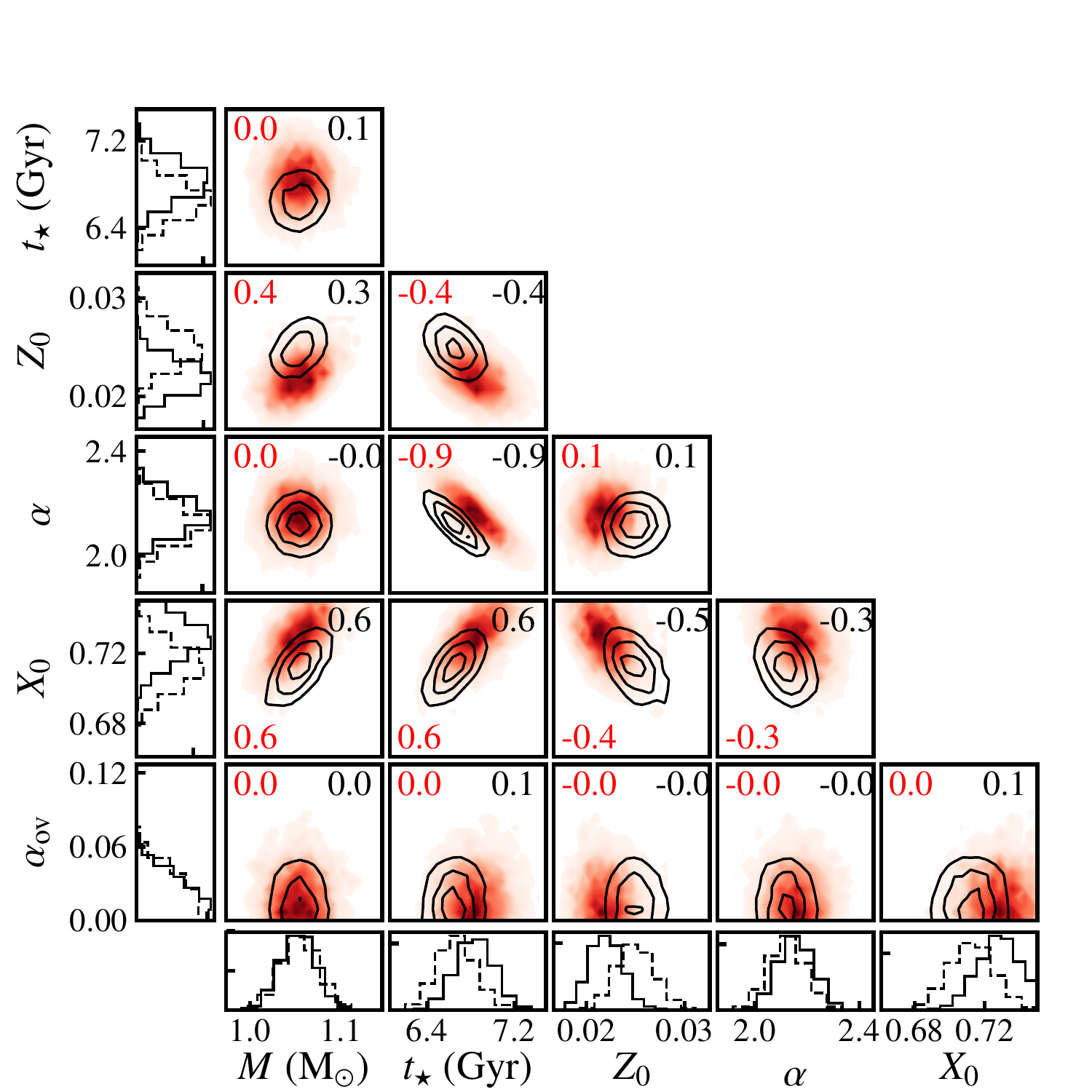}
   \caption{Marginal densities for the stellar parameters $M$, $\stage$, $X_0$, $Z_0$, $\alpha$, and $\alphaov$ of 16~Cyg~A (left) and B (right). The central panels show the joint marginal densities of the paired parameters. Individual marginal densities are plotted in the side panels. The red shaded areas in the central panels and the full lines in the side panels represent posterior densities for models with penetrative overshoot and helium diffusion (\#5 in Tables~\ref{tab:params_16cygA} and \ref{tab:params_16cygB}). The black contours in the central panels and the dashed lines in the side panels represent posterior densities for models with penetrative and metal diffusion (\#6 in Tables~\ref{tab:params_16cygA} and \ref{tab:params_16cygB}). The numbers in each panel give the Pearson correlation coefficient for the two variables. The values in red correspond to the red shaded densities and the values in black to the black contours.
   }
         \label{fig:joint-16CygAB-stellar-overshoot}
\end{figure*}

 Microscopic diffusion of chemical abundances plays an important role in the characterisation of the solar sound speed profile, which in turn relates to the oscillation frequencies\citep{Bahcall95,JCD96,Richard96}. Evidence of gravitational settling of heavy elements has also been found in more massive stars \citep[e.g][]{Richard01}. The \emph{Kepler} data may help to understand the magnitude of the bias induced by neglecting diffusion for all or some elements. In theory, if we consider particles settling in an hydrogen background, then all heavier elements should be included. However, treatments of diffusion in the literature are somewhat inconsistent. First of all, diffusion has not been systematically included. This has the advantage of helping to set the initial hydrogen-mass fraction and metallicity, at least for stars up until the first dredge-up, since their surface ratio does not evolve and correspond to the observed metallicity. Diffusion is sometimes also excluded from stellar models due to numerical issues. For ASTEC difficulties may arise when treating diffusion of elements heavier than helium when the star has a convective core \citep{JCD08a}. This is why only helium diffusion was considered in \citet{Metcalfe15} and \citet{Creevey17}. Sometimes diffusion is neglected above a certain stellar mass \citep{SA17,Creevey17}. The argument given is that the short diffusive timescales corresponding to shallow convective envelopes would deplete the upper layers of elements heavier than hydrogen too fast. This actually reflects the incompleteness of most stellar models that do not include {\lq}radiative levitation{\rq}, which becomes an efficient competing phenomena at intermediate masses \citep[e.g.][]{Richard02}. 16 Cyg A and B present the advantage of having low enough mass (even when accounting for the uncertainties on this parameter) so that no convective core ever develops \citep[this is also due to their almost solar metallicity; see e.g. ][]{Bazot12,Bazot16}.

  I ran two additional simulations for each star: the first does not include diffusion and the second takes into account settling of helium and heavy elements. The resulting two-dimensional and one-dimensional posterior marginal densities are shown in Fig.~\ref{fig:joint-16CygAB-stellar-diffusion}. It should first be noticed that this does not change the magnitude of the estimated uncertainties for any parameter. When diffusion is not included, the estimated age is higher than in the baseline case and the mixing-length parameter is lower. Within their 68.3\% credible intervals, all estimates agree albeit marginally. The estimated $X_0$ and $Z_0$ are lower and higher, respectively, but they remain closer to those of the baseline case. When diffusion of metal is included, the trends for $X_0$ and $Z_0$ are similar, but deviate more from the baseline case. The estimate of $X_0$, which now decreases by more than $1\sigma$ and $Z_0$ increases by $\sim$15\% (1.8$\sigma$ from baseline case) for 16 Cyg A and $\sim$11\% (1.3$\sigma$ from baseline case).  This is a well-known effect resulting from the depletion of the external convective envelope of its metals. In order to reproduce the observed surface metallicity, the average value $Z_0$ needs to increase. However, contrary to the non-diffusion case, the estimated age this time decreases and the mixing-length parameter does not change.

\begin{figure*}[h!]
   \centering
   \includegraphics[width=.45\textwidth]{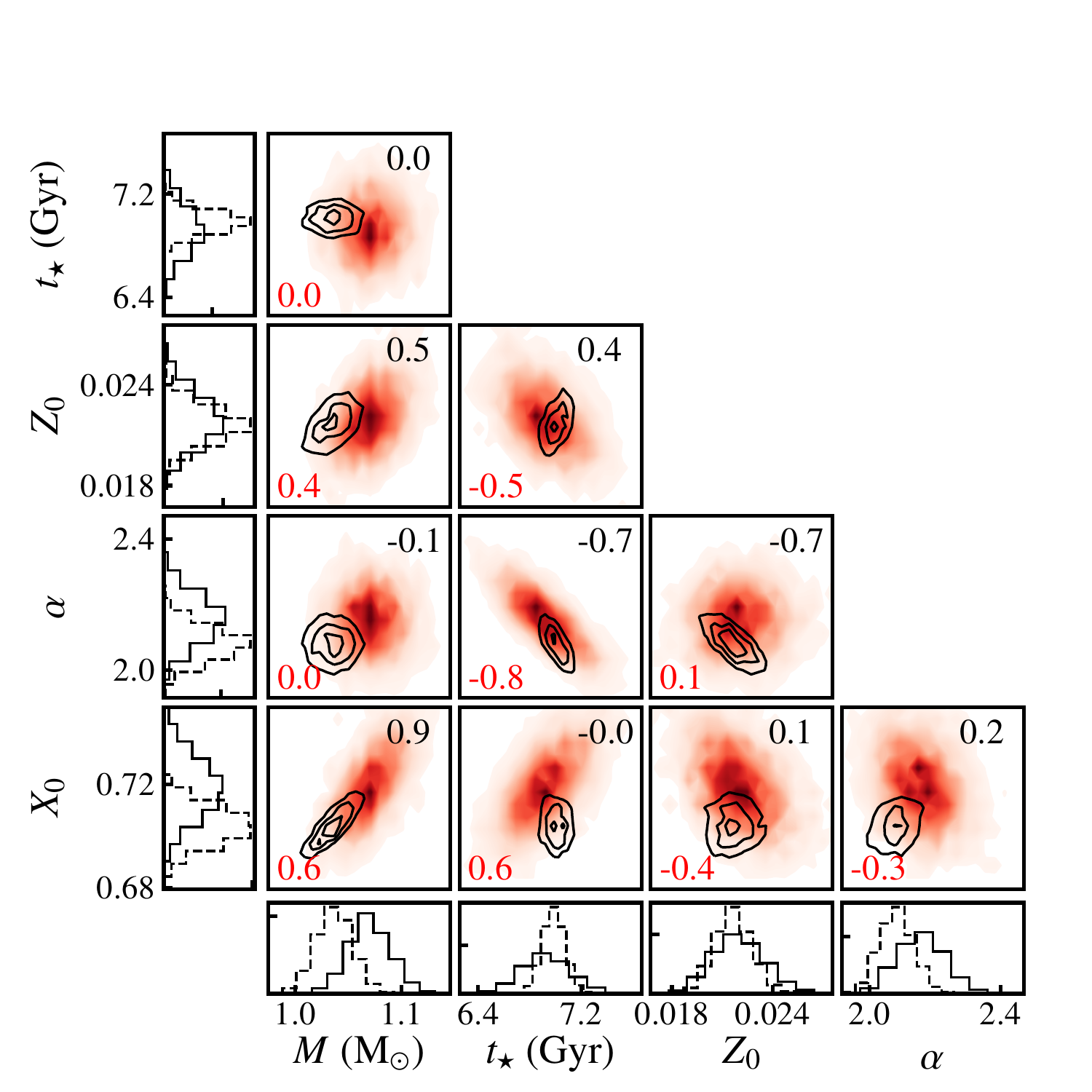}
   \includegraphics[width=.45\textwidth]{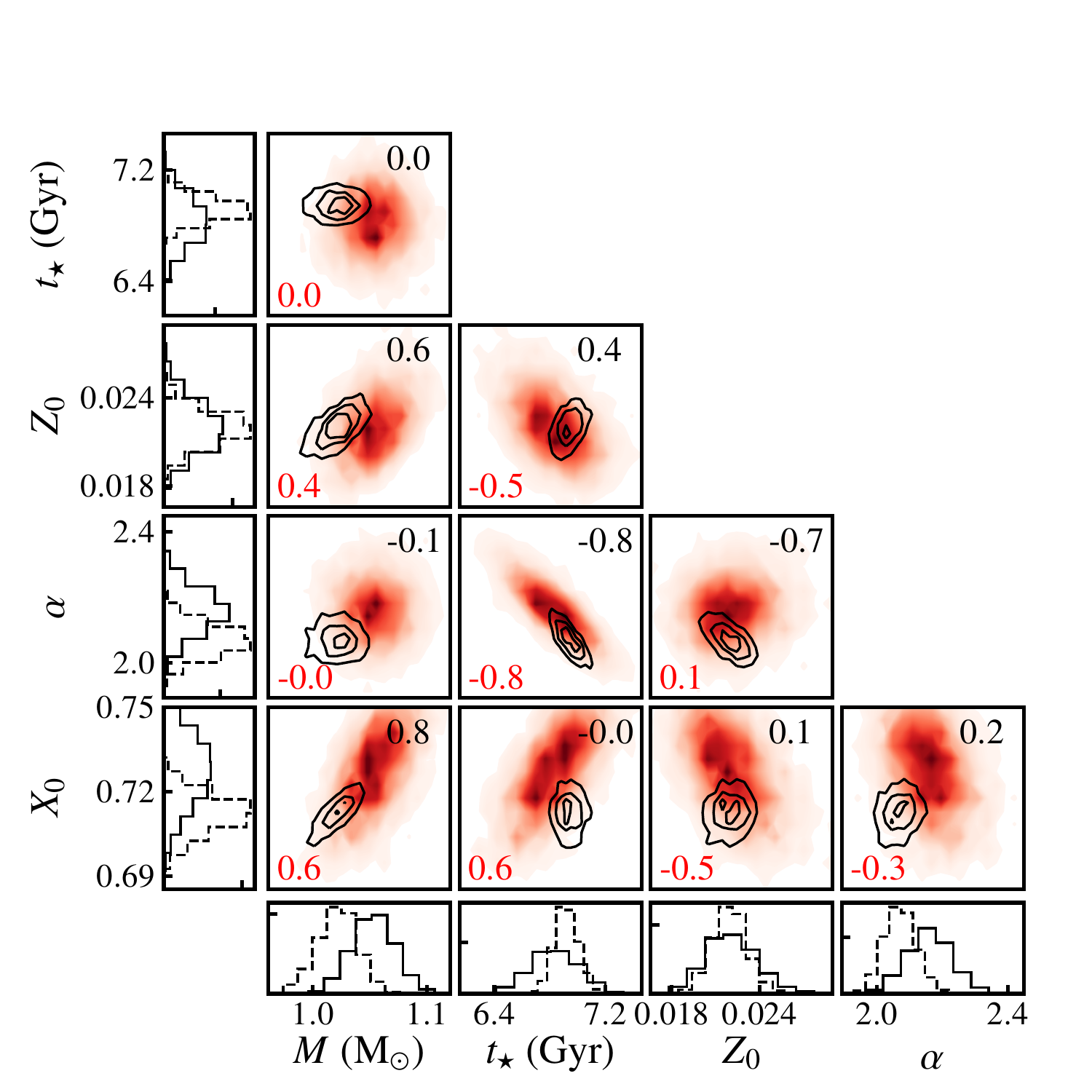}
   \caption{Marginal densities for the stellar parameters $M$, $\stage$, $X_0$, $Z_0$, and $\alpha$ of 16~Cyg~A (left) and B (right). The central panels show the joint marginal densities of the paired parameters. Individual marginal densities are plotted in the side panels.  The red shaded areas in the central panels and the full lines in the side panels represent posterior densities for the baseline case. The black contours  in the central panels and the dashed lines in the side panels represent posterior densities when $\rzoz$ is used as a seismic constraint instead of $\rzo$. 
   }
         \label{fig:joint-16CygAB-stellar-diagnostics}
\end{figure*}
  This gives us some insight into potential biases caused by the details of microscopic diffusion in stellar models. To that effect, I now consider that the most accurate physical model is that including microscopic diffusion for the metals. Then, provided that  the asteroseismic data are similar in quality to the \emph{Kepler} time series for 16 Cyg A and B and that the star is in a physical state close enough to the solar state, we could expect that neglecting diffusion implies higher estimated ages and lower mixing-length parameters. The initial metallicity and hydrogen-mass fraction may also be marginally affected, respectively, decreasing and increasing. On the other hand, including helium diffusion does not so strongly impact the estimates for the age and the mixing-length parameter, but induces higher deviation in the initial chemical composition. The strongest bias is on the age when diffusion is omitted. To conclude, it should be noted that even moderate bias may become important when attempting to perform statistical studies such as those seen in stellar open cluster analysis or Galactic archaeology. This may cause the averages used for the quantities of interest to converged towards a wrong value.

\subsection{Overshooting}\label{sect:overshoot} 

  Overshooting is a long-standing problem in stellar physics. Stellar models use mixing-length theory to model convection. It is a local theory in which convective motions are expected to stop at the boundary of the unstable envelope defined by the Schwarzschild criteria, that is when the radiative temperature gradient equals the adiabatic temperature gradient. It has long been recognised that convective movements are likely to persist beyond this limit \citep{Veronis63,Moore67}. It is not clear however whether downward convective flows are strong enough to thermalise the subadiabatically stratified layers below the convective boundary. This is the case when the P\'eclet is high enough; such a situation is often referred to as penetration \citep{Zahn91}. This is opposed to chemical mixing, when overshooting only homogenises abundances without affecting the temperature gradient. In this study I follow the recommendation of \citet{Viallet15}, based on a semi-quantitative analysis, which suggests using penetration to model overshoot below the convective envelope. The depth of the overshoot/penetration zone is set using
\begin{equation}\label{eq:povs}
\ell_{\mathrm{ov}}  = \alphaov H_p,
\end{equation}
where $\ell_{\mathrm{ov}}$ is the size of the overshoot layer and $H_p$ the pressure scale height at the boundary of the convective envelope.

There have been a few estimates of the overshoot parameter for 16 Cyg A \& B. In \citet{SA17}, six stellar evolution codes out of seven include overshoot. However, the overshoot parameter has been fixed to calibrated values. A notable exception to this lack of published estimates is the work of \citet{Bellinger16} who obtained values for the overshoot parameter of $0.07\pm0.03$ and $0.11\pm0.03$ for 16 Cyg A \& B, respectively. Their prescription for overshooting differs from the present one in that chemical mixing (and not penetration) is achieved through a diffusive process \citep{Herwig00}.

 I ran two additional simulations for each star to estimate $\alphaov$. In the first simulation only helium settles. In the second, diffusion of metals is taken into account. The idea is that the diffusion of heavy elements affects the layers immediately below the convective zone and that it may impact overshoot, which is expected to occur in the same region. The corresponding PDFs of the stellar parameters are shown in Fig.~\ref{fig:joint-16CygAB-stellar-overshoot} and their estimates are given in Tables \ref{tab:params_16cygA} and \ref{tab:params_16cygB}. The first obvious result is that $\alphaov$ is poorly constrained. The marginal densities for $\alphaov$, contrary to the other parameters, are not Gaussian, but closer to a gamma distribution. From the marginal MAP estimates I obtain uncertainties above 100\%. This is in sharp contrast with the results from \citet{Bellinger16} who estimated uncertainties of the order of 40\%.  It is not clear if this discrepancy is due to the difference in the estimation strategy or in the physical modelling. It is also noteworthy that $\alphaov$ does not correlate with any other stellar parameters. There is therefore no bias to be found in the other stellar parameters for the inclusion of penetrative overshoot. We see that $\alphaov$ is marginally larger than when only helium is diffusing. However, the uncertainties remain extremely large and the two prescriptions cannot be distinguished.

In order to estimate $\alphaov$ for both overshoot and penetration, I ran an additional simulation for each star. Besides the inclusion of penetrative convection, the physical models are identical to the baseline case. The corresponding PDFs of the stellar parameters are shown in Fig.~\ref{fig:joint-16CygAB-stellar-overshoot} and their estimates are given in Tables \ref{tab:params_16cygA} and \ref{tab:params_16cygB}. The first obvious result is that $\alphaov$ is poorly constrained. The marginal densities for $\alphaov$, contrary to the other parameters, are not Gaussian, but closer to a gamma distribution. From the marginal MAP estimates I obtain uncertainties above 100\%. This is in sharp contrast with the results from \citet{Bellinger16} who estimated uncertainties of the order of 40\%.  It is not clear if this discrepancy is due to the difference in the estimation strategy or in the physical modelling. It is also noteworthy that $\alphaov$ does not correlate with any other stellar parameters. There is therefore no bias to be found in the other stellar parameters for the inclusion of penetrative overshoot.

The extent of the overshotting region in the Sun has been measured by several authors. For instance \citet{Monteiro94} and \citet{JCD95} have used solar models including penetration to reproduce the signatures that sharp transitions in the sound speed profile induce in the oscillation frequencies. These authors estimate a depth for the solar overshoot region in an approximate range 0.07$H_p$ -- 0.1$H_p$. When helium diffusion only is included, the upper limits of the 68.3\% credible intervals for $\alphaov$ are below these solar values for both 16 Cyg and B. When metal diffusion is included, the 68.3\% credible intervals for $\alphaov$ overlaps marginally with the range given for the solar value. 

Numerical investigations have also been used to evaluate the characteristics of stellar overshooting. They are obviously limited in that they they assume Prandtl numbers that are too high and Rayleigh numbers that are too low with respect to a typical stellar interior. These numerical invesigations nevertheless provide interesting insight into the general behaviour of stellar convection. Many studies have been published considering either fluids in the Boussinesq approximation, the anelastic approximation, or fully compressible fluids. Broadly speaking two types of results emerge depending on whether they are the models are two- or three-dimensional. Two-dimensional simulations \citep{Hurlburt86,Hurlburt94,Rogers05,Rogers06} often exhibit penetrative overshoot. On the other hand, in three-dimensional simulations \citep{Singh95,Brummel02,Korre19} penetration is not observed but rather chemical mixing occurs down to significant depth. This is most likely due to a larger density of plume-like structure in two-dimensional than in three-dimensional  simulations. Our results, even though they do not address chemical mixing, are consistent with the picture of a very small to non-existent penetration region.

\subsection{Using $r_{01}$/$r_{10}$ ratios}\label{sect:ratios}

The final MCMC runs for both 16 Cyg A and B aim at assessing the effect of using $\rzoz$ instead of $\rzo$. The obvious significant result is that the estimates of the masses depart from those of all other runs. The discrepancy is significant, i.e. of the order of 3\% for both stars, which corresponds to $\sim$1.5$\sigma$ from the baseline case. The corresponding posterior densities are shown in Fig.~\ref{fig:joint-16CygAB-stellar-diagnostics}. In light of the criticism expressed in \citet{Roxburgh18}, this discrepancy can be interpreted as a bias with respect to the {\lq}`correct{'\rq} value of the parameter, which according to the previous discussions are likely to be around 1.07~{\msol} and 1.05~{\msol} for 16 Cyg A and B, respectively. The use of $\rzoz$ also biases the results towards lower $X_0$ and $\alpha$ values and, marginally, towards higher values of $\stage$. The uncertainties on the age are also biased when using $\rzoz$ as seismic constraints. The estimates in this case are twice as small as those obtained using $\rzo$.

The effect uncovered in this work happens to be quite subtle. Indeed, it has already been pointed out that the results for the baseline case are in good agreement with those of \citet{Metcalfe15} who used $\rzoz$ and not $\rzo$ as done in this case. This can be explained by the fact that they neglected the non-diagonal terms in $\Sigmav$. To test this interpretation, I use MCMC simulations for both 16 Cyg A an B using $\rzoz$ but computing the likelihood (\ref{eq:likelihood}) using only the diagonal terms of $\Sigmav$. I then find mass estimates in agreement with the baseline case.

The general picture that emerges from these results is that decent models for 16 Cyg A and B may be found at masses around 1.04~{\msol} and 1.02~\msol, respectively. Their emergence as local or global minima strongly depends on the precise computation of the likelihood, in particular on the non-diagonal terms of $\Sigmav$. A good indicator of this behaviour is the value the argument of the exponential in Eq.~(\ref{eq:likelihood}), that is the $\chi^2$ values, of the best models for the baseline case and the case constrained using $\rzoz$. In the former case, it is indeed larger for the model at 1.07~\msol, while in the latter it is larger for the 1.04 {\msol} model. This behaviour is reversed again if the non-diagonal terms in $\Sigmav$ are set to zero. This shows how delicate this problem can be. There are competing effects and, in the case of 16 Cyg A and B, the bias induced by over-fitting is compensated by the approximation that consists in neglecting correlations between frequency ratios. However, there is no indication that this behaviour can be straightforwardly generalised to other stars. Therefore, the present results strengthen the more general claim of \citet{Roxburgh18}, outlining the need to properly take into account the correlations between seismic indicator and using $\rzo$ (or $r_{10}$) rather than $\rzoz$ ratios.

\section{Conclusions}

In this study, I used Bayesian statistics to derive robust estimates and uncertainties for the physical parameters of the solar analogues 16 Cyg A and B. Using a statistical method independent from the previous studies on this stars, I obtain a precision on the mass of the order of 4\% and on the age of the order of 6\%. This is in fair agreement with most of the published estimates. I outline the need to use the proper seismic diagnostics to avoid biases on the mass, initial chemical composition, and mixing-length parameter (and, marginally, the age), and their uncertainties. I also pointed out the changes induced in the estimates of the initial chemical composition induced by changes in the abundances ratios and microscopic diffusion. The latter, limiting the comparison to the abundance ratio of \citet{Grevesse93} and \citet{Grevesse98}, are negligible. The effects of microscopic diffusion are much more important. They may cause biases in the estimates in the range 7\% -- 8\%. They may also introduce some biases in the estimated initial metallicity. Considering penetrative overshoot below the convective envelope, I find that the sub-adiabatic region can be thermalised only in a very shallow layer. Including penetrative overshoot does not affect the estimates of the other parameters. The results presented in this study complete those previously obtained on 16 Cyg A and B, which are excellent benchmark of what can be achieved using precise asteroseismic data. The future PLATO mission will allow us to observe many more Sun-like stars. The present results aim at helping to prepare their modelling.

\begin{acknowledgements}
I thank the referee for his/her helpful comments, which really help improving this article. I thank S. Hannestad for providing him access to the Grendel cluster at DCSC/AU of which important use has been made during this work. I also thank T. Metcalfe for answering many questions on his previous work, this was extremely helpful in constructing this analysis. This material is based upon work partly supported by the NYU Abu Dhabi Institute under grant G1502. 
\end{acknowledgements}

\bibliography{ref}

\begin{thebibliography}{79}
\expandafter\ifx\csname natexlab\endcsname\relax\def\natexlab#1{#1}\fi

\bibitem[{{Angulo} {et~al.}(1999){Angulo}, {Arnould}, {Rayet}, {Descouvemont},
  {Baye}, {Leclercq-Willain}, {Coc}, {Barhoumi}, {Aguer}, {Rolfs}, {Kunz},
  {Hammer}, {Mayer}, {Paradellis}, {Kossionides}, {Chronidou}, {Spyrou},
  {degl'Innocenti}, {Fiorentini}, {Ricci}, {Zavatarelli}, {Providencia},
  {Wolters}, {Soares}, {Grama}, {Rahighi}, {Shotter}, \& {Lamehi
  Rachti}}]{Angulo99}
{Angulo}, C., {Arnould}, M., {Rayet}, M., {et~al.} 1999, Nuclear Physics A,
  656, 3

\bibitem[{{Angulo} {et~al.}(2005){Angulo}, {Champagne}, \&
  {Trautvetter}}]{Angulo05}
{Angulo}, C., {Champagne}, A.~E., \& {Trautvetter}, H.-P. 2005, Nuclear Physics
  A, 758, 391

\bibitem[{{Antia} \& {Basu}(2011)}]{Antia11}
{Antia}, H.~M. \& {Basu}, S. 2011, Journal of Physics Conference Series, 271,
  012034

\bibitem[{{Asplund} {et~al.}(2005){Asplund}, {Grevesse}, \&
  {Sauval}}]{Asplund05}
{Asplund}, M., {Grevesse}, N., \& {Sauval}, A.~J. 2005, in Astronomical Society
  of the Pacific Conference Series, Vol. 336, Cosmic Abundances as Records of
  Stellar Evolution and Nucleosynthesis, ed. T.~G. {Barnes}, III \& F.~N.
  {Bash}, 25

\bibitem[{{Asplund} {et~al.}(2009){Asplund}, {Grevesse}, {Sauval}, \&
  {Scott}}]{Asplund09}
{Asplund}, M., {Grevesse}, N., {Sauval}, A.~J., \& {Scott}, P. 2009, \araa, 47,
  481

\bibitem[{{Baglin} {et~al.}(2009){Baglin}, {Auvergne}, {Barge}, {Deleuil},
  {Michel}, \& {CoRoT Exoplanet Science Team}}]{Baglin09}
{Baglin}, A., {Auvergne}, M., {Barge}, P., {et~al.} 2009, in IAU Symposium,
  Vol. 253, Transiting Planets, ed. F.~{Pont}, D.~{Sasselov}, \& M.~J.
  {Holman}, 71--81

\bibitem[{{Bahcall} {et~al.}(1995){Bahcall}, {Pinsonneault}, \&
  {Wasserburg}}]{Bahcall95}
{Bahcall}, J.~N., {Pinsonneault}, M.~H., \& {Wasserburg}, G.~J. 1995, Reviews
  of Modern Physics, 67, 781

\bibitem[{{Basu}(2016)}]{Basu16}
{Basu}, S. 2016, Living Reviews in Solar Physics, 13, 2

\bibitem[{{Basu} \& {Antia}(2008)}]{Basu08}
{Basu}, S. \& {Antia}, H.~M. 2008, \physrep, 457, 217

\bibitem[{{Bazot}(2013)}]{Bazot13}
{Bazot}, M. 2013, in EAS Publications Series, Vol.~63, EAS Publications Series,
  ed. G.~{Alecian}, Y.~{Lebreton}, O.~{Richard}, \& G.~{Vauclair}, 105--114

\bibitem[{{Bazot} {et~al.}(2019){Bazot}, {Benomar}, {Christensen-Dalsgaard},
  {Gizon}, {Hanasoge}, {Nielsen}, {Petit}, \& {Sreenivasan}}]{Bazot19}
{Bazot}, M., {Benomar}, O., {Christensen-Dalsgaard}, J., {et~al.} 2019, arXiv
  e-prints

\bibitem[{{Bazot} {et~al.}(2007){Bazot}, {Bouchy}, {Kjeldsen}, {Charpinet},
  {Laymand}, \& {Vauclair}}]{Bazot07}
{Bazot}, M., {Bouchy}, F., {Kjeldsen}, H., {et~al.} 2007, \aap, 470, 295

\bibitem[{{Bazot} {et~al.}(2008){Bazot}, {Bourguignon}, \&
  {Christensen-Dalsgaard}}]{Bazot08}
{Bazot}, M., {Bourguignon}, S., \& {Christensen-Dalsgaard}, J. 2008, Memorie
  della Societa Astronomica Italiana, 79, 660

\bibitem[{{Bazot} {et~al.}(2012){Bazot}, {Campante}, {Chaplin}, {Carfantan},
  {Bedding}, {Dumusque}, {Broomhall}, {Petit}, {Th{\'e}ado}, {Van Grootel},
  {Arentoft}, {Castro}, {Christensen-Dalsgaard}, {do Nascimento}, {Dintrans},
  {Kjeldsen}, {Monteiro}, {Santos}, {Sousa}, \& {Vauclair}}]{Bazot12}
{Bazot}, M., {Campante}, T.~L., {Chaplin}, W.~J., {et~al.} 2012, \aap, 544,
  A106

\bibitem[{{Bazot} {et~al.}(2016){Bazot}, {Christensen-Dalsgaard}, {Gizon}, \&
  {Benomar}}]{Bazot16}
{Bazot}, M., {Christensen-Dalsgaard}, J., {Gizon}, L., \& {Benomar}, O. 2016,
  \mnras, 460, 1254

\bibitem[{{Bazot} {et~al.}(2018){Bazot}, {Creevey}, {Christensen-Dalsgaard}, \&
  {Mel{\'e}ndez}}]{Bazot18}
{Bazot}, M., {Creevey}, O., {Christensen-Dalsgaard}, J., \& {Mel{\'e}ndez}, J.
  2018, \aap, 619, A172

\bibitem[{{Bedding} {et~al.}(2001){Bedding}, {Butler}, {Kjeldsen}, {Baldry},
  {O'Toole}, {Tinney}, {Marcy}, {Kienzle}, \& {Carrier}}]{Bedding01}
{Bedding}, T.~R., {Butler}, R.~P., {Kjeldsen}, H., {et~al.} 2001, \apjl, 549,
  L105

\bibitem[{{Bedding} {et~al.}(2007){Bedding}, {Kjeldsen}, {Arentoft}, {Bouchy},
  {Brandbyge}, {Brewer}, {Butler}, {Christensen-Dalsgaard}, {Dall}, {Frandsen},
  {Karoff}, {Kiss}, {Monteiro}, {Pijpers}, {Teixeira}, {Tinney}, {Baldry},
  {Carrier}, \& {O'Toole}}]{Bedding07}
{Bedding}, T.~R., {Kjeldsen}, H., {Arentoft}, T., {et~al.} 2007, \apj, 663,
  1315

\bibitem[{{Bellinger} {et~al.}(2016){Bellinger}, {Angelou}, {Hekker}, {Basu},
  {Ball}, \& {Guggenberger}}]{Bellinger16}
{Bellinger}, E.~P., {Angelou}, G.~C., {Hekker}, S., {et~al.} 2016, \apj, 830,
  31

\bibitem[{{Bellinger} {et~al.}(2017){Bellinger}, {Basu}, {Hekker}, \&
  {Ball}}]{Bellinger17}
{Bellinger}, E.~P., {Basu}, S., {Hekker}, S., \& {Ball}, W.~H. 2017, \apj, 851,
  80

\bibitem[{Berger \& Berger(1985)}]{Berger85}
Berger, J.~O. \& Berger, J.~O. 1985, Statistical decision theory and Bayesian
  analysis (New York: Springer-Verlag)

\bibitem[{Birnbaum(1962)}]{Birnbaum62}
Birnbaum, A. 1962, Journal of the American Statistical Association, 57, 269

\bibitem[{{B{\"o}hm-Vitense}(1958)}]{BV58}
{B{\"o}hm-Vitense}, E. 1958, Zeitschrift f\"ur Astrophysik, 46, 108

\bibitem[{{Borucki} {et~al.}(2010){Borucki}, {Koch}, {Basri}, {Batalha},
  {Brown}, {Caldwell}, {Caldwell}, {Christensen-Dalsgaard}, {Cochran},
  {DeVore}, {Dunham}, {Dupree}, {Gautier}, {Geary}, {Gilliland}, {Gould},
  {Howell}, {Jenkins}, {Kjeldsen}, {Kondo}, {Latham}, {Lissauer}, {Marcy},
  {Meibom}, {Monet}, {Morrison}, {Sasselov}, \& {Tarter}}]{Borucki10}
{Borucki}, W.~J., {Koch}, D., {Basri}, G., {et~al.} 2010, in Bulletin of the
  American Astronomical Society, Vol.~42, American Astronomical Society Meeting
  Abstracts \#215, 215

\bibitem[{{Bouchy} {et~al.}(2005){Bouchy}, {Bazot}, {Santos}, {Vauclair}, \&
  {Sosnowska}}]{Bouchy05}
{Bouchy}, F., {Bazot}, M., {Santos}, N.~C., {Vauclair}, S., \& {Sosnowska}, D.
  2005, \aap, 440, 609

\bibitem[{{Bouchy} \& {Carrier}(2001)}]{Bouchy01}
{Bouchy}, F. \& {Carrier}, F. 2001, \aap, 374, L5

\bibitem[{Brooks \& Gelman(1998)}]{Brooks98}
Brooks, S.~P. \& Gelman, A. 1998, Journal of Computational and Graphical
  Statistics, 7, 434

\bibitem[{{Brown} {et~al.}(1994){Brown}, {Christensen-Dalsgaard},
  {Weibel-Mihalas}, \& {Gilliland}}]{Brown94}
{Brown}, T.~M., {Christensen-Dalsgaard}, J., {Weibel-Mihalas}, B., \&
  {Gilliland}, R.~L. 1994, \apj, 427, 1013

\bibitem[{{Brummell} {et~al.}(2002){Brummell}, {Clune}, \&
  {Toomre}}]{Brummel02}
{Brummell}, N.~H., {Clune}, T.~L., \& {Toomre}, J. 2002, \apj, 570, 825

\bibitem[{{Castro} {et~al.}(2007){Castro}, {Vauclair}, \& {Richard}}]{Castro07}
{Castro}, M., {Vauclair}, S., \& {Richard}, O. 2007, \aap, 463, 755

\bibitem[{{Christensen-Dalsgaard}({2008a})}]{JCD08a}
{Christensen-Dalsgaard}, J. {2008a}, \apss, 316, 13

\bibitem[{{Christensen-Dalsgaard}({2008b})}]{JCD08b}
{Christensen-Dalsgaard}, J. {2008b}, \apss, 316, 113

\bibitem[{{Christensen-Dalsgaard} {et~al.}(1996){Christensen-Dalsgaard},
  {Dappen}, {Ajukov}, {Anderson}, {Antia}, {Basu}, {Baturin}, {Berthomieu},
  {Chaboyer}, {Chitre}, {Cox}, {Demarque}, {Donatowicz}, {Dziembowski},
  {Gabriel}, {Gough}, {Guenther}, {Guzik}, {Harvey}, {Hill}, {Houdek},
  {Iglesias}, {Kosovichev}, {Leibacher}, {Morel}, {Proffitt}, {Provost},
  {Reiter}, {Rhodes}, {Rogers}, {Roxburgh}, {Thompson}, \& {Ulrich}}]{JCD96}
{Christensen-Dalsgaard}, J., {Dappen}, W., {Ajukov}, S.~V., {et~al.} 1996,
  Science, 272, 1286

\bibitem[{{Christensen-Dalsgaard} {et~al.}(1995){Christensen-Dalsgaard},
  {Monteiro}, \& {Thompson}}]{JCD95}
{Christensen-Dalsgaard}, J., {Monteiro}, M. J.~P.~F.~G., \& {Thompson}, M.~J.
  1995, \mnras, 276, 283

\bibitem[{Clayton(1968)}]{Clayton68}
Clayton, D. 1968, Principles of stellar evolution and nucleosynthesis
  (University of Chicago Press)

\bibitem[{{Creevey} {et~al.}(2017){Creevey}, {Metcalfe}, {Schultheis},
  {Salabert}, {Bazot}, {Th{\'e}venin}, {Mathur}, {Xu}, \&
  {Garc{\'{\i}}a}}]{Creevey17}
{Creevey}, O.~L., {Metcalfe}, T.~S., {Schultheis}, M., {et~al.} 2017, \aap,
  601, A67

\bibitem[{{Creevey} {et~al.}(2007){Creevey}, {Monteiro}, {Metcalfe}, {Brown},
  {Jim{\'e}nez-Reyes}, \& {Belmonte}}]{Creevey07}
{Creevey}, O.~L., {Monteiro}, M.~J.~P.~F.~G., {Metcalfe}, T.~S., {et~al.} 2007,
  \apj, 659, 616

\bibitem[{{Cunha} {et~al.}(2007){Cunha}, {Aerts}, {Christensen-Dalsgaard},
  {Baglin}, {Bigot}, {Brown}, {Catala}, {Creevey}, {Domiciano de Souza},
  {Eggenberger}, {Garcia}, {Grundahl}, {Kervella}, {Kurtz}, {Mathias},
  {Miglio}, {Monteiro}, {Perrin}, {Pijpers}, {Pourbaix}, {Quirrenbach},
  {Rousselet-Perraut}, {Teixeira}, {Th{\'e}venin}, \& {Thompson}}]{Cunha07}
{Cunha}, M.~S., {Aerts}, C., {Christensen-Dalsgaard}, J., {et~al.} 2007, \aapr,
  14, 217

\bibitem[{{Davies} {et~al.}(2015){Davies}, {Chaplin}, {Farr}, {Garc{\'{\i}}a},
  {Lund}, {Mathis}, {Metcalfe}, {Appourchaux}, {Basu}, {Benomar}, {Campante},
  {Ceillier}, {Elsworth}, {Handberg}, {Salabert}, \& {Stello}}]{Davies15}
{Davies}, G.~R., {Chaplin}, W.~J., {Farr}, W.~M., {et~al.} 2015, \mnras, 446,
  2959

\bibitem[{Gelman \& Rubin(1992)}]{Gelman92}
Gelman, A. \& Rubin, D.~B. 1992, Statist. Sci., 7, 457

\bibitem[{{Gough}(2012)}]{Gough12}
{Gough}, D.~O. 2012, in Astronomical Society of the Pacific Conference Series,
  Vol. 462, Progress in Solar/Stellar Physics with Helio- and Asteroseismology,
  ed. H.~{Shibahashi}, M.~{Takata}, \& A.~E. {Lynas-Gray}, 429

\bibitem[{{Grevesse} \& {Noels}(1993)}]{Grevesse93}
{Grevesse}, N. \& {Noels}, A. 1993, in Origin and Evolution of the Elements,
  ed. N.~{Prantzos}, E.~{Vangioni-Flam}, \& M.~{Casse}, 15--25

\bibitem[{{Grevesse} \& {Sauval}(1998)}]{Grevesse98}
{Grevesse}, N. \& {Sauval}, A.~J. 1998, \ssr, 85, 161

\bibitem[{{Guzik} {et~al.}(2006){Guzik}, {Watson}, \& {Cox}}]{Guzik06}
{Guzik}, J.~A., {Watson}, L.~S., \& {Cox}, A.~N. 2006, \memsai, 77, 389

\bibitem[{Haario {et~al.}(2001)Haario, Saksman, \& Tamminen}]{Haario01}
Haario, H., Saksman, E., \& Tamminen, J. 2001, Bernoulli, 7, 223

\bibitem[{{Herwig}(2000)}]{Herwig00}
{Herwig}, F. 2000, \aap, 360, 952

\bibitem[{{Hurlburt} {et~al.}(1986){Hurlburt}, {Toomre}, \&
  {Massaguer}}]{Hurlburt86}
{Hurlburt}, N.~E., {Toomre}, J., \& {Massaguer}, J.~M. 1986, \apj, 311, 563

\bibitem[{{Hurlburt} {et~al.}(1994){Hurlburt}, {Toomre}, {Massaguer}, \&
  {Zahn}}]{Hurlburt94}
{Hurlburt}, N.~E., {Toomre}, J., {Massaguer}, J.~M., \& {Zahn}, J.-P. 1994,
  \apj, 421, 245

\bibitem[{{Iglesias} \& {Rogers}(1996)}]{Iglesias96}
{Iglesias}, C.~A. \& {Rogers}, F.~J. 1996, \apj, 464, 943

\bibitem[{{Kjeldsen} {et~al.}(2005){Kjeldsen}, {Bedding}, {Butler},
  {Christensen-Dalsgaard}, {Kiss}, {McCarthy}, {Marcy}, {Tinney}, \&
  {Wright}}]{Bedding05}
{Kjeldsen}, H., {Bedding}, T.~R., {Butler}, R.~P., {et~al.} 2005, \apj, 635,
  1281

\bibitem[{{Kjeldsen} {et~al.}(2008){Kjeldsen}, {Bedding}, \&
  {Christensen-Dalsgaard}}]{Kjeldsen08}
{Kjeldsen}, H., {Bedding}, T.~R., \& {Christensen-Dalsgaard}, J. 2008, \apjl,
  683, L175

\bibitem[{{Korre} {et~al.}(2019){Korre}, {Garaud}, \& {Brummell}}]{Korre19}
{Korre}, L., {Garaud}, P., \& {Brummell}, N.~H. 2019, \mnras, 484, 1220

\bibitem[{{Metcalfe} {et~al.}(2012){Metcalfe}, {Chaplin}, {Appourchaux},
  {Garc{\'{\i}}a}, {Basu}, {Brand{\~a}o}, {Creevey}, {Deheuvels}, {Do{\v g}an},
  {Eggenberger}, {Karoff}, {Miglio}, {Stello}, {Y{\i}ld{\i}z}, {{\c C}elik},
  {Antia}, {Benomar}, {Howe}, {R{\'e}gulo}, {Salabert}, {Stahn}, {Bedding},
  {Davies}, {Elsworth}, {Gizon}, {Hekker}, {Mathur}, {Mosser}, {Bryson},
  {Still}, {Christensen-Dalsgaard}, {Gilliland}, {Kawaler}, {Kjeldsen},
  {Ibrahim}, {Klaus}, \& {Li}}]{Metcalfe12}
{Metcalfe}, T.~S., {Chaplin}, W.~J., {Appourchaux}, T., {et~al.} 2012, \apjl,
  748, L10

\bibitem[{{Metcalfe} {et~al.}(2009){Metcalfe}, {Creevey}, \&
  {Christensen-Dalsgaard}}]{Metcalfe09}
{Metcalfe}, T.~S., {Creevey}, O.~L., \& {Christensen-Dalsgaard}, J. 2009, \apj,
  699, 373

\bibitem[{{Metcalfe} {et~al.}(2015){Metcalfe}, {Creevey}, \&
  {Davies}}]{Metcalfe15}
{Metcalfe}, T.~S., {Creevey}, O.~L., \& {Davies}, G.~R. 2015, \apjl, 811, L37

\bibitem[{{Metcalfe} {et~al.}(2014){Metcalfe}, {Creevey}, {Do{\u g}an},
  {Mathur}, {Xu}, {Bedding}, {Chaplin}, {Christensen-Dalsgaard}, {Karoff},
  {Trampedach}, {Benomar}, {Brown}, {Buzasi}, {Campante}, {{\c C}elik},
  {Cunha}, {Davies}, {Deheuvels}, {Derekas}, {Di Mauro}, {Garc{\'{\i}}a},
  {Guzik}, {Howe}, {MacGregor}, {Mazumdar}, {Montalb{\'a}n}, {Monteiro},
  {Salabert}, {Serenelli}, {Stello}, {Ste{\c e}\'slicki}, {Suran},
  {Y{\i}ld{\i}z}, {Aksoy}, {Elsworth}, {Gruberbauer}, {Guenther}, {Lebreton},
  {Molaverdikhani}, {Pricopi}, {Simoniello}, \& {White}}]{Metcalfe14}
{Metcalfe}, T.~S., {Creevey}, O.~L., {Do{\u g}an}, G., {et~al.} 2014, \apjs,
  214, 27

\bibitem[{{Michaud} \& {Proffitt}(1993)}]{Michaud93}
{Michaud}, G. \& {Proffitt}, C.~R. 1993, in Astronomical Society of the Pacific
  Conference Series, Vol.~40, IAU Colloq. 137: Inside the Stars, ed. W.~W.
  {Weiss} \& A.~{Baglin}, 246--259

\bibitem[{{Monteiro} {et~al.}(1994){Monteiro}, {Christensen-Dalsgaard}, \&
  {Thompson}}]{Monteiro94}
{Monteiro}, M.~J.~P.~F.~G., {Christensen-Dalsgaard}, J., \& {Thompson}, M.~J.
  1994, \aap, 283, 247

\bibitem[{{Moore}(1967)}]{Moore67}
{Moore}, D.~W. 1967, in IAU Symposium, Vol.~28, Aerodynamic Phenomena in
  Stellar Atmospheres, ed. R.~N. {Thomas}, 405

\bibitem[{{Porto de Mello} {et~al.}(2014){Porto de Mello}, {da Silva}, {da
  Silva}, \& {de Nader}}]{PdM14}
{Porto de Mello}, G.~F., {da Silva}, R., {da Silva}, L., \& {de Nader}, R.~V.
  2014, \aap, 563, A52

\bibitem[{{Richard} {et~al.}(2001){Richard}, {Michaud}, \&
  {Richer}}]{Richard01}
{Richard}, O., {Michaud}, G., \& {Richer}, J. 2001, \apj, 558, 377

\bibitem[{{Richard} {et~al.}(2002){Richard}, {Michaud}, {Richer}, {Turcotte},
  {Turck-Chi{\`e}ze}, \& {VandenBerg}}]{Richard02}
{Richard}, O., {Michaud}, G., {Richer}, J., {et~al.} 2002, \apj, 568, 979

\bibitem[{{Richard} {et~al.}(1996){Richard}, {Vauclair}, {Charbonnel}, \&
  {Dziembowski}}]{Richard96}
{Richard}, O., {Vauclair}, S., {Charbonnel}, C., \& {Dziembowski}, W.~A. 1996,
  \aap, 312, 1000

\bibitem[{Robert(2007)}]{Robert07}
Robert, C. 2007, The Bayesian Choice: From Decision-Theoretic Foundations to
  Computational Implementation, Springer Texts in Statistics (Springer New
  York)

\bibitem[{Robert \& Casella(2005)}]{Robert05}
Robert, C.~P. \& Casella, G. 2005, Monte Carlo Statistical Methods (Springer
  Texts in Statistics) (Secaucus, NJ, USA: Springer-Verlag New York, Inc.)

\bibitem[{{Rogers} \& {Nayfonov}(2002)}]{OPAL02}
{Rogers}, F.~J. \& {Nayfonov}, A. 2002, \apj, 576, 1064

\bibitem[{{Rogers} \& {Glatzmaier}(2005)}]{Rogers05}
{Rogers}, T.~M. \& {Glatzmaier}, G.~A. 2005, \apj, 620, 432

\bibitem[{{Rogers} {et~al.}(2006){Rogers}, {Glatzmaier}, \& {Jones}}]{Rogers06}
{Rogers}, T.~M., {Glatzmaier}, G.~A., \& {Jones}, C.~A. 2006, \apj, 653, 765

\bibitem[{{Roxburgh}(2017)}]{Roxburgh17}
{Roxburgh}, I.~W. 2017, \aap, 604, A42

\bibitem[{{Roxburgh}(2018)}]{Roxburgh18}
{Roxburgh}, I.~W. 2018, arXiv e-prints

\bibitem[{{Roxburgh} \& {Vorontsov}(2003)}]{RV03}
{Roxburgh}, I.~W. \& {Vorontsov}, S.~V. 2003, \aap, 411, 215

\bibitem[{{Silva Aguirre} {et~al.}(2013){Silva Aguirre}, {Basu}, {Brand{\~a}o},
  {Christensen-Dalsgaard}, {Deheuvels}, {Do{\u g}an}, {Metcalfe}, {Serenelli},
  {Ballot}, {Chaplin}, {Cunha}, {Weiss}, {Appourchaux}, {Casagrande},
  {Cassisi}, {Creevey}, {Garc{\'{\i}}a}, {Lebreton}, {Noels}, {Sousa},
  {Stello}, {White}, {Kawaler}, \& {Kjeldsen}}]{SA13}
{Silva Aguirre}, V., {Basu}, S., {Brand{\~a}o}, I.~M., {et~al.} 2013, \apj,
  769, 141

\bibitem[{{Silva Aguirre} {et~al.}(2017){Silva Aguirre}, {Lund}, {Antia},
  {Ball}, {Basu}, {Christensen-Dalsgaard}, {Lebreton}, {Reese}, {Verma},
  {Casagrande}, {Justesen}, {Mosumgaard}, {Chaplin}, {Bedding}, {Davies},
  {Handberg}, {Houdek}, {Huber}, {Kjeldsen}, {Latham}, {White}, {Coelho},
  {Miglio}, \& {Rendle}}]{SA17}
{Silva Aguirre}, V., {Lund}, M.~N., {Antia}, H.~M., {et~al.} 2017, \apj, 835,
  173

\bibitem[{{Singh} {et~al.}(1995){Singh}, {Roxburgh}, \& {Chan}}]{Singh95}
{Singh}, H.~P., {Roxburgh}, I.~W., \& {Chan}, K.~L. 1995, \aap, 295, 703

\bibitem[{{Vauclair} {et~al.}(2008){Vauclair}, {Laymand}, {Bouchy}, {Vauclair},
  {Hui Bon Hoa}, {Charpinet}, \& {Bazot}}]{Vauclair08}
{Vauclair}, S., {Laymand}, M., {Bouchy}, F., {et~al.} 2008, \aap, 482, L5

\bibitem[{{Veronis}(1963)}]{Veronis63}
{Veronis}, G. 1963, \apj, 137, 641

\bibitem[{{Viallet} {et~al.}(2015){Viallet}, {Meakin}, {Prat}, \&
  {Arnett}}]{Viallet15}
{Viallet}, M., {Meakin}, C., {Prat}, V., \& {Arnett}, D. 2015, \aap, 580, A61

\bibitem[{von Toussaint(2011)}]{vonToussaint11}
von Toussaint, U. 2011, Rev. Mod. Phys., 83, 943

\bibitem[{{Zahn}(1991)}]{Zahn91}
{Zahn}, J.-P. 1991, \aap, 252, 179

\end{thebibliography}

\appendix

\section{Convergence of the MCMC algorithm}\label{app:conv}

\begin{figure}[h!]
   \centering
   \includegraphics[width=.45\textwidth]{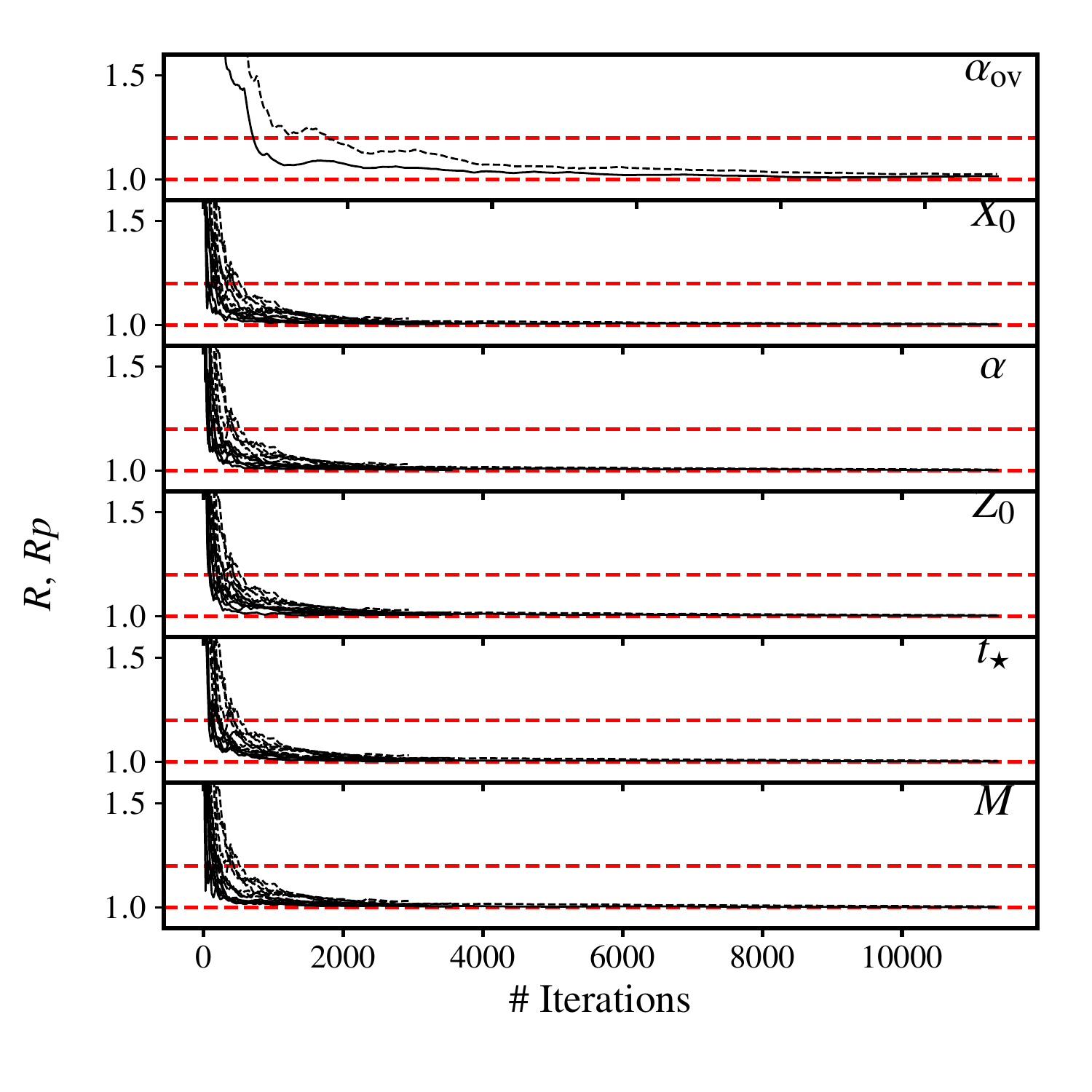}
   \caption{Pseudo-scale ($R$, full lines) and multivariate pseudo-scale factors ($R_p$, dashed line) for all cases presented in Table~\ref{tab:params_16cygA} 16 Cyg A. The dashed horizontal red lines denote the 1 and 1.2 values.}
         \label{fig:16CygA-convergence}
   \end{figure}
\begin{figure}[h!]
   \centering
   \includegraphics[width=.45\textwidth]{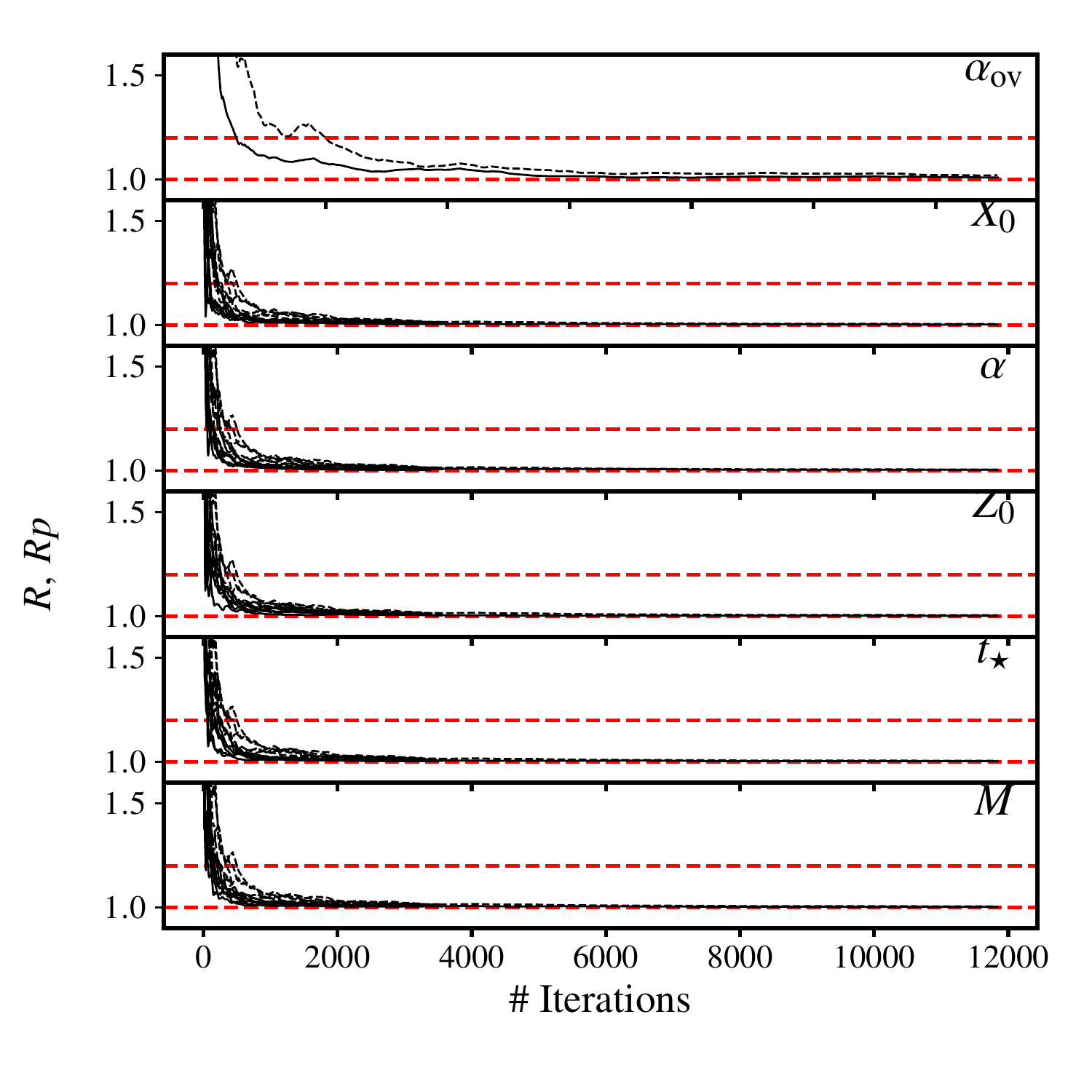}
   \caption{Pseudo-scale ($R$, full lines) and multivariate pseudo-scale factors ($R_p$, dashed line) for all cases presented in Table~\ref{tab:params_16cygB} 16 Cyg B. The dashed horizontal red lines indicate the 1 and 1.2 values.}
         \label{fig:16CygB-convergence}
   \end{figure}

In this Appendix, I provide assessments on the convergence of the MCMC simulations. In Figs. ~\ref{fig:16CygA-convergence} and \ref{fig:16CygB-convergence}, I show the pseudo-scale \citep{Gelman92} and multivariate pseudo-scale \citep{Brooks98} criteria. The general rule-of-thumb regarding convergence is to have these criteria below $\sim1.2$. In the limit of infinitely many iterations, they can be expected to tend to one. All the runs are well behaved with respect to these convergence diagnostics. The multivariate pseudo-scale factor is always superior to pseudo-scale factor, which is expected, the latter converging to one faster than the former.

In theory, a single chain run for a very long time, as well as several shorter chains, would sample the target density. However, given the large computational cost of stellar models, the current set-up represents a real gain in that samples generated from the same stationary distribution can be merged to get larger samples, Furthermore, such tests as those performed in this work are not available with one single chain, hence representing an operational advantage.

\end{document}